\documentclass[12pt]{iopart}
\usepackage[english]{babel}
\usepackage[utf8x]{inputenc}
\usepackage[T1]{fontenc}
\usepackage{braket}
\usepackage{graphicx}
\usepackage{multirow}
\usepackage{booktabs}
\usepackage{colortbl}
\usepackage{appendix}
\usepackage{xcolor}
\usepackage{indentfirst}
\usepackage[export]{adjustbox}
\usepackage{subcaption}
\usepackage{caption}
\usepackage{lipsum}
\usepackage{cite}
\expandafter\let\csname equation*\endcsname\relax
\expandafter\let\csname endequation*\endcsname\relax
\usepackage{amsmath}
\usepackage{amssymb}
\usepackage{marginnote}
\usepackage{graphicx}
\usepackage{dirtytalk}
\usepackage[colorinlistoftodos]{todonotes}
\usepackage[colorlinks=true, allcolors=blue]{hyperref}

\begin{document}

\title[]{Insights on Entanglement Entropy in $1+1$ Dimensional Causal Sets}

\author{Th\'eo Keseman, Hans J. Muneesamy, and  Yasaman K. Yazdi}
\address{Theoretical Physics Group, Blackett Laboratory, Imperial College London, %

SW7 2AZ, UK}
\ead{theo.keseman17@imperial.ac.uk, hans.muneesamy17@imperial.ac.uk, ykouchek@imperial.ac.uk}
\vspace{10pt}

\begin{abstract}
Entanglement entropy in causal sets offers a fundamentally covariant characterisation of quantum field degrees of freedom. A known result in this context is that the degrees of freedom consist of a number of contributions that have continuum-like analogues, in addition to a number of contributions that do not. The latter exhibit features below the discreteness scale and are excluded from the entanglement entropy using a ``truncation scheme''. This truncation is necessary to recover the standard spatial area law of entanglement entropy. In this paper we build on previous work on the entanglement entropy of a massless scalar field on a causal set approximated by a $1+1$D causal diamond in Minkowski spacetime. We present new insights into the truncated contributions, including evidence that they behave as fluctuations and encode features specific to a particular causal set sprinkling. We extend previous results in the massless theory to include R\'enyi entropies and include new results for the massive theory. We also discuss the implications of our work for the treatment of entanglement entropy in causal sets in more general settings.
\end{abstract}

\section{Introduction}
An important open question that quantum gravity aims to answer is: what are the microscopic degrees of freedom or states that describe black hole entropy? We know that the classical Bekenstein-Hawking black hole entropy is finite and proportional to the area of the event horizon \cite{ PhysRevD.7.2333, PhysRevD.9.3292, cmp/1103899181}. The proportionality constant is also known and is $1/4$ when some fundamental constants are set to $1$. Whatever the microscopic nature of this entropy is and however many states there are then, they too must be finite, and yield an entropy equal to $1/4$ of the event horizon area in the same fundamental units. A similar statement applies to other horizons, such as cosmological ones, with analogous thermodynamic properties.

Entanglement entropy of quantum fields is one promising candidate microscopic source of this entropy. In fact, while entanglement entropy has a wide array of applications today, it was originally conceived of to study this very question of the microscopic nature of black hole entropy \cite{sorkin1983entropy}. From its earliest days \cite{sorkin1983entropy, 22} it was noticed that entanglement entropy also generically scales as the area of the boundary of the entangling subregion, be this the event horizon in a black hole spacetime or more generally some surface dividing the entire quantum system into two complementary subsystems. The proportionality constant in these so-called spatial ``area laws" is not fixed, and depends on the details of the  UV physics. Another crucial observation made early on in these studies was that the entanglement entropy diverges in the absence of a UV cutoff. Thus a UV cutoff is necessary to obtain a finite entanglement entropy and for it to have a chance to be the correct microscopic picture of black hole entropy. 

If this UV cutoff is furnished by a theory of quantum gravity such as causal set theory, this would bring us closer to understanding the horizon degrees of freedom fundamentally. Moreover, if we wish to understand the $1/4$ proportionality constant in addition to the area  scaling, then our fundamental theory (as well as our entanglement entropy definition) must be covariant. The study of entanglement entropy in causal sets allows one to do exactly this. Causal set theory is equipped with a covariant discreteness scale which serves as the UV cutoff. Furthermore, a definition due to Sorkin \cite{18} of entanglement entropy in terms of spacetime correlation functions, makes possible the use of this covariant cutoff in counting quantum field degrees of freedom. Hence it is of interest to study this subject further.

The most well understood calculation of 
entanglement entropy in causal set theory is in the context of a Gaussian massless scalar field in a $1+1$D causal set approximated by a causal diamond in Minkowski spacetime \cite{20}. The entanglement was considered between the field restriction to a smaller subdiamond and its complement in the larger one in which the global pure state was defined. Some challenges were met along the way, but with insights from analytic calculations in the continuum \cite{1,71,59,19, 8}, these were overcome and the expected area laws of the entanglement entropy were obtained. The challenges stemmed from a number of unexpected, and dominating, contributions to the entanglement entropy. 

A key ingredient in the calculation, are the eigenfunctions and eigenvalues of the Pauli-Jordan function or spacetime commutator. Previous work in the $1+1$D continuum diamond \cite{1, 59} has shown that the eigenfunctions are (approximately) linear combinations of plane waves and the eigenvalues are inversely proportional to the wavenumbers which have known values following a power law. 
In the causal set we also see discrete analogues of these plane wave eigenfunctions and their eigenvalues, all the way down to wavelengths near the discreteness scale. The contributions do not stop here, however. Instead, we also observe numerous eigenfunctions that do not look like plane waves and have eigenvalues that do not follow a power law. This latter set of contributions is poorly understood and it is an open question whether or not they have any physical significance. As first discovered in \cite{20}, this  family of contributions must be excluded from the entanglement entropy calculations in order to recover the conventional area laws. The exclusion is done via a ``truncation scheme'', whereby these solutions are eliminated at two stages of the calculation. Without the truncations, a spacetime volume law is obtained.

Attempts have been made to extend the results of \cite{20} to more general settings such as higher dimensions, nonlocal field theories, disjoint regions, and spacetimes with curvature \cite{57, Surya_2021, duffy2021entanglement}. These attempts have had various degrees of success. What these studies have shown is that a better understanding of the extra contributions in the causal set, either through closer investigations in the causal set or through further analytic studies in the continuum, is necessary to motivate unique generalisations of the truncation scheme. 

In this work, we return to the best understood case of the $1+1$D causal diamond, with the intention to gain further insights from the causal set. We study more closely the eigenfunctions that do not have continuum-like counterparts and ask whether they are in some sense fluctuations below the discreteness scale. We find  evidence in favor of this. Specifically, we consider averages of the eigenfunctions over many sprinklings. We find that the continuum-like contributions have persistent features while the others do not. The details of this are  presented in Section \ref{fluct}. 

The plan of the remainder of the paper is as follows. We begin by reviewing the spacetime entanglement entropy formalism, as well as the expected scaling laws in $1+1$D that we will compare to, in Section \ref{EE}. In Section \ref{mless} we reproduce the results of the massless theory, and include new results on R\'enyi entropies. We then discuss the massive theory in Section \ref{mive}. The extension to the massive theory is nontrivial, as we lack analytic results there. Upon close inspection of the Pauli-Jordan function eigenvalues and eigenfunctions in this case, we find that the knowledge from the massless theory is adequate to employ a meaningful truncation here. This is encouraging and has implications for more general extensions of these and similar results. These implications are discussed in Section \ref{summary}.

\section{Entanglement Entropy}
\label{EE}
Many different techniques exist for computing  entanglement entropy. 
Below we review the spacetime formulation of entanglement entropy for a Gaussian scalar field theory, defined in \cite{18}, which we use throughout this paper. This formulation is special because it is in terms of \emph{spacetime} correlation functions and hence allows for a covariant treatment of the degrees of freedom and UV cutoff.

\subsection{Spacetime Definition of Entropy}
 We will consider a Gaussian scalar field theory. Therefore, everything, including the entanglement entropy, can be determined from the two-point correlation function or Wightman function.\footnote{Interestingly, it has been shown that the Wightman function is also enough to determine the entanglement entropy in non-Gaussian and interacting theories up to first order in perturbation theory \cite{68}.}
The entropy associated to the scalar field in a spacetime region $\mathcal{R}$ (where $\mathcal{R}$ can be a full spacetime or causal set, or a subset of it) is 

\begin{equation}\label{see}
S=\sum_\lambda \lambda\, \text{ln} |\lambda|,
\end{equation}
where $\lambda$ is a solution to
\begin{equation}\label{eigeigeig}
    W\, v = i \lambda \Delta v
\end{equation}
under the condition
\begin{equation}\label{condition}
    \Delta v \neq 0.
\end{equation}

$W$ is the Wightman function $W(x,x')=\bra0\phi(x)\phi(x')\ket{0}$ and $i\Delta$, which is equal to one half of the imaginary part of $W$, is the Pauli-Jordan function or spacetime commutator  $i\Delta(x,x')=[\phi(x),\phi(x')]$, where $x,x'\in \mathcal{R}$. That the expression \eqref{see} agrees with the standard $S(\rho)=-\text{Tr}\, \rho\, \text{ln} \rho$ can be readily seen from its derivation via the replica trick, given in \cite{68}, which we review in  \ref{app:replica}.

 Analogously, the Rényi entropy of order $\alpha$ is
\begin{equation}\label{ftr}
    S^{(\alpha)}=\frac{-1}{1-\alpha}\sum_\lambda \text{ln}(\lambda^\alpha - (\lambda-1)^\alpha),
\end{equation}
where each term in the sum \eqref{ftr} accounts for a pair of eigenvalues $\lambda$ and $(1-\lambda)$.

If the global state is taken to be the SJ state, $W=W_{SJ}$, we have a pure state and the entropy vanishes. The SJ Wightman function\footnote{We will review the SJ state in greater detail in the next section.} is defined to be the restriction of $i\Delta$ to its positive eigenspace, $W_{SJ}\equiv \text{Pos}(i\Delta)$. This results in all the eigenvalues in \eqref{see} being either $\lambda=0$ or $\lambda=1$ and the entropy vanishing, as mentioned. If, however,  $W$ and $i\Delta$ are restricted to subsets of the full degrees of freedom or subregions within $\mathcal{R}$, we can obtain a non-zero entropy. In particular, this occurs if the subregion has a non-zero causal complement, and the entropy in this case can be regarded as the entanglement entropy between the subregion and its causal complement.

As mentioned above, entanglement entropy diverges in the absence of a UV cutoff. Therefore to obtain a meaningful entanglement entropy, we need a UV regulator or fundamental cutoff. The kind of cutoff that can be applied is also dependent on the manner in which the entanglement entropy is calculated. For example, when the calculation is done on a spatial hypersurface (as it often is), common choices of cutoffs include a minimum spatial distance from the entangling boundary or a spatial lattice spacing. The formulation presented in this subsection allows for the implementation of a spacetime cutoff. $W$ and $i\Delta$ are both spacetime functions, though infinite dimensional in the continuum. One possible way to implement a covariant spacetime cutoff, as was done in \cite{19}, is to expand $W$ and $i\Delta$ in the eigenbasis of $i\Delta$\footnote{The closure of the image of the $i\Delta$, or its eigenfunctions with non-zero eigenvalues, span the full solution space of the Klein Gordon equation \cite{17}.} up to a minimum eigenvalue which can be related to a minimum wavelength or maximum wavenumber and defined as the cutoff. In the causal set, as we will see below, we are automatically equipped with the discreteness scale as our cutoff, and both $W$ and $i\Delta$ are finite dimensional matrices. 

\subsection{Scaling Laws in $1+1$D}
We will study both a massless and massive scalar field theory in $1+1$D. We will compare our results for the  entanglement and R\'enyi entropies with known scaling laws that have been derived and numerically verified in various places in the literature. Below we summarise the relevant scalings for later reference. Most of these scaling laws have been studied in settings where the entanglement entropy is calculated on a spatial hypersurface. These results can be compared to those in our spacetime formulation, if the spacetime regions considered are domains of dependence (or causally convex subsets of the domains of dependence, containing the Cauchy surface) of the spatial regions. It is worth highlighting that the spacetime formulation reviewed in the previous subsection does not require for the spacetime or subregions to be domains of dependence of spatial Cauchy hypersurfaces. This merely facilitates things such as comparisons to  conventional results.

The entanglement entropy of a massless scalar field confined to a spatial interval of length $\Tilde{\ell}$ within a larger spatial interval of length $\Tilde{L}>>\Tilde{\ell}$, is \cite{Calabrese_2009}

\begin{equation}
S\sim \frac{1}{3}\text{ln}\bigg(\frac{\Tilde{\ell}}{a}\bigg)+c_1
\label{area}
\end{equation}
where $a$ is the UV cutoff, and $c_1$ is a non-universal constant. This scaling is for the case where the subinterval $\Tilde{\ell}$ has two boundaries across which the entanglement occurs and in the limit of small UV cutoff $a$.

Similarly, the Rényi entanglement entropy of order $\alpha$ is \cite{Calabrese_2009}
\begin{equation}
\label{cardy_renyi}
S^{(\alpha)}\sim \frac{1}{6}\bigg(1+\frac{1}{\alpha}\bigg)\text{ln}\bigg(\frac{\Tilde{\ell}}{a}\bigg)+c_\alpha
\end{equation}
where $c_\alpha$ are non-universal constants.

The entanglement entropy of a massive scalar field with mass $m$, where $a\ll1/m<\Tilde{\ell}$ is \cite{Calabrese_2009}
\begin{equation}
\label{cardy_massive}
S\sim -\frac{1}{3}\text{ln}(ma)+\Tilde{c}_1,
\end{equation}
where $\Tilde{c}_1$ is once again a non-universal constant.
Note that these logarithmic scalings are consistent with the general arguments that lead to area laws for the entanglement entropy \cite{Chandran_2016}. The scalings \eqref{area} and \eqref{cardy_renyi} have been obtained in the continuum using the spacetime formulation of the previous subsection in \cite{19}. The scaling \eqref{area} has been obtained in the causal set using the same formulation, in \cite{20,8}. Below we present results confirming all of these scalings in the causal set. 

\section{Massless Scalar Field Theory}
\label{mless}
A Gaussian scalar field theory on a causal set can be set up using the Sorkin-Johnston (SJ) prescription \cite{1,17,29}. This prescription uses the entire spacetime and yields a unique and covariant Wightman function, the SJ Wightman function $W_{SJ}$. This Wightman function can then be used in \eqref{eigeigeig} to obtain the entanglement entropy. The starting point of this prescription is the retarded Green function\footnote{The SJ prescription defines a unique $W_{SJ}$ in globally hyperbolic spacetimes (or causal sets approximated by globally hyperbolic spacetimes), since only these spacetimes possess a unique retarded Green function.}, which for a scalar field with mass $m$ in D-dimensional Minkowski spacetime satisfies
\begin{equation}\label{eq:2.9}
(\Box - m^2) G_R^{(D)}(x,\,x') = -\delta^{(D)}(x-x')
\end{equation}
where $x=(t,\Vec{x})$ and $G_R^{(D)}(x,\,x') =0$ unless $x'$ causally precedes $x$, denoted by $x' \prec x$ \cite{66,13}. For a massless scalar field in $1+1$D Minkowski spacetime, this Green function has the simple form

\begin{equation}\label{eq:2.10}
G_R^{(2)}(x,\,x') = \frac{1}{2}\theta(t-t')\theta(-\tau^2)
\end{equation}
where $\theta$ is the Heaviside step function and $\tau=\sqrt{(\Delta \vec{x})^2-(\Delta t)^2}$ is the proper time. In other words, this Green function is only non-zero if $x'$ is in the past lightcone of $x$ and takes the constant value of $1/2$ if this is the case. The causal matrix $C$ can be used to define a discrete analogue of this Green function in the causal set. The causal set retarded Green function is

\begin{equation}\label{cgr}
G_{R,xx'} = \frac{1}{2}C_{x'x},
\end{equation}
where $C$ is the causal matrix
\begin{equation}
\label{cmatrix}
C_{xx'}=\left\{\begin{matrix}1
 & \text{for $x\prec x'$}\\ 0
 & \text{otherwise}.
\end{matrix}\right.
\end{equation}
The causal set Green function \eqref{cgr} agrees identically with the continuum Green function \eqref{eq:2.10} evaluated at the causal set elements. The spacetime field commutator, or Pauli-Jordan function, is then given by
\begin{equation}\label{eq:2.13}
i\Delta = i(G_{R} - G_A)
\end{equation}
where $G_A$, the advanced Green function, is simply the transpose of $G_R$. $i\Delta$ is anti-symmetric and Hermitian and has real non-zero pairs of eigenvalues, $\pm \Tilde{\lambda}_i$ and eigenvectors given by
\begin{flalign}
\begin{split}
&i\Delta u_i=\Tilde{\lambda_i} u_i,\\
&i\Delta v_i=-\Tilde{\lambda_i} v_i,\\
&i\Delta w_k=0
\end{split}
\end{flalign}
where $\Tilde{\lambda_i} >0$. 

\begin{figure}[h!]
    \centering
    \includegraphics[width=0.65\linewidth]{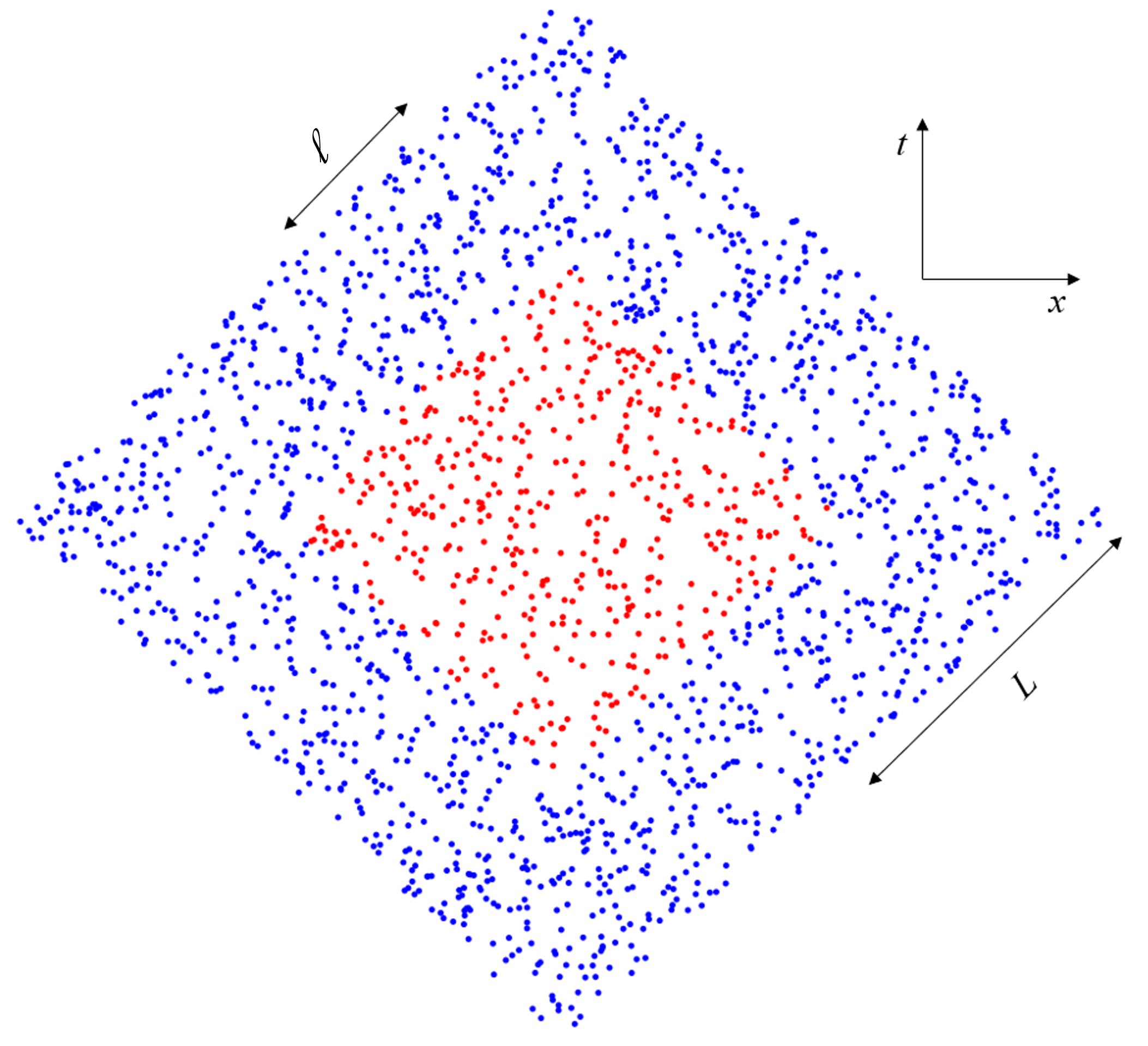}
    \caption{Causal set of a smaller causal diamond (with side length $2\ell$) nested within a larger causal diamond (with side length $2L$) of 2000 elements generated via a Poisson sprinkling.}
    \label{sprinkling2}
\end{figure}

$i\Delta$ can then be expanded in its non-zero eigenbasis as 
\begin{equation}\label{eq:2.14}
i\Delta = \sum_{i=1}\Tilde{\lambda}_i u_i u_i^\dagger - \Tilde{\lambda}_i v_i v_i^\dagger.
\end{equation}
Restricting only to the positive eigenspace of $i\Delta$, the SJ Wightman function is
\begin{equation}\label{eq:2.15}
W_{SJ} \equiv \text{Pos}(i\Delta) = \sum_{i=1}\Tilde{\lambda}_i u_i u_i^\dagger.
\end{equation}
Once we have $W$, we have all we need to compute the entropy using \eqref{see} and \eqref{eigeigeig}. We will first review the results of \cite{20} and the truncation scheme, before presenting new results. Throughout this paper, we will work with a causal set approximated by a causal diamond in $1+1$D Minkowski spacetime, and we will consider the entanglement entropy arising from the restriction of $W_{SJ}$ to a smaller subdiamond concentric to the larger one. These regions are shown in Figure \ref{sprinkling2}, with the subdiamond in red. The properties of $W_{SJ}$ and the eigenspectrum of $i\Delta$ in these causal diamonds has been well studied \cite{71,59,19}. In the massless theory, working with a compact region such as a causal diamond cures the IR divergence otherwise present in $1+1$D. In addition, the causal diamond is globally hyperbolic (hence there will be a unique $G_R$).  

In Minkowski light-cone coordinates $u=\frac{t+x}{\sqrt{2}}$ and $v=\frac{t-x}{\sqrt{2}}$, with $u,v \in [-L,L]$, the Pauli-Jordan function in the causal diamond is
\begin{equation}
    \Delta(u,v;u',v')=-\frac{1}{2}(\theta(u-u')+\theta(v-v')-1).
\end{equation}
Its eigenfunctions $f$ in the continuum satisfy 
\begin{equation}\label{eigvalprob}
    \int_{-L}^{L}du\int_{-L}^{L}dv\, i \Delta(u-u',v-v') f(u,v) = \Tilde{\lambda} f(u',v').
\end{equation}
The full set of functions with non-zero eigenvalues satisfying \eqref{eigvalprob} is known and form a two-set family of eigenfunctions, $f_k$ and $g_k$. These are
\begin{flalign}\label{eigmassless}
\begin{split}
&f_k(u,\,v):=e^{-iku}-e^{-ikv} \text{ with } k=\frac{n \pi}{L}, n=\pm\,1,\pm\,2,...\\
&g_k(u,\,v):=e^{-iku}+e^{-ikv}-2\text{cos}(kL) \text{ with } k\in \mathcal{K},
\end{split}
\end{flalign}
where $\mathcal{K}=\{k \in \mathbb{R}| \text{tan}(kL)=2kL \text{ and } k \neq 0\}$ \cite{1,20}. The eigenvalues for both sets are 
\begin{equation}
\label{lamt}
\Tilde{\lambda}_k=\frac{L}{k}.    
\end{equation}

\begin{figure}[h!]
    \centering
    \includegraphics[width=0.7\linewidth]{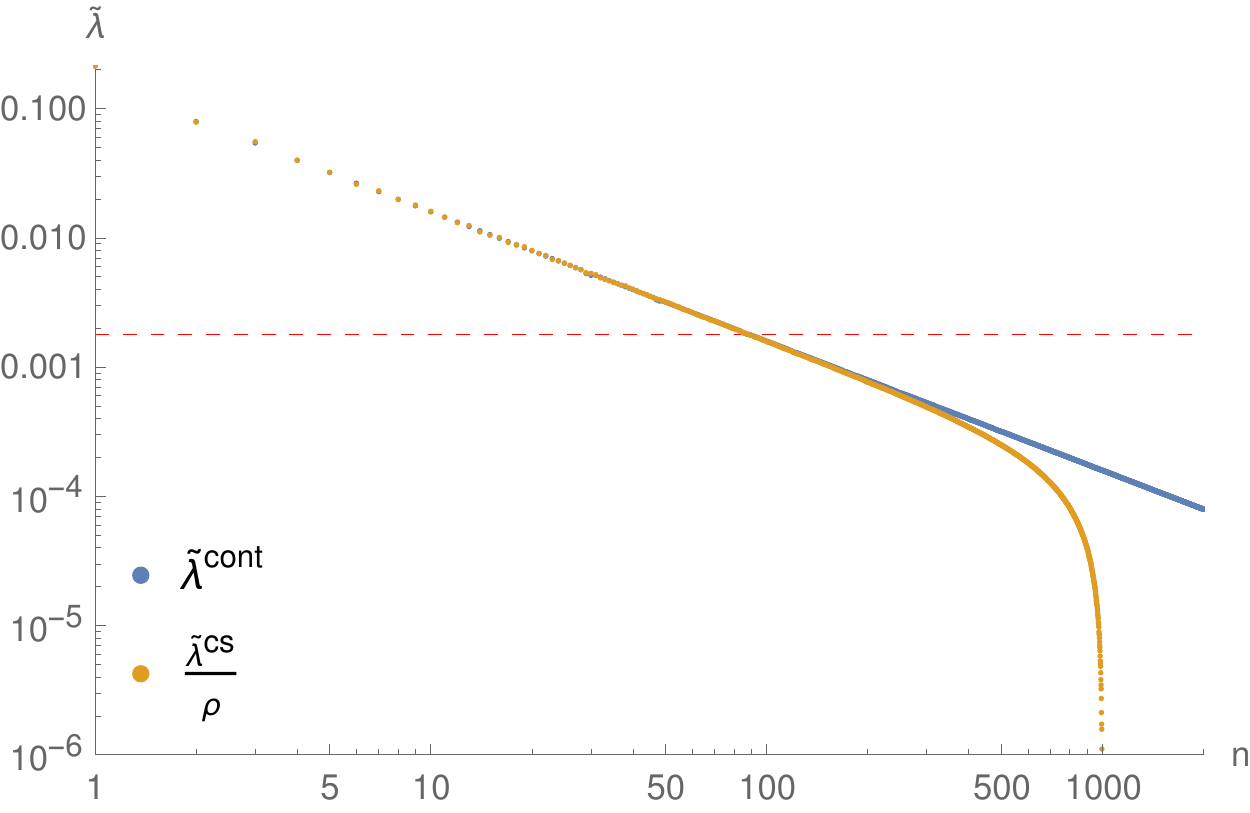}
    \caption{Comparison (on a log-log scale) of the positive spectrum of $i\Delta$ in the continuum and the causal set for a causal set of 2000 elements and $L=\frac{1}{2}$. The horizontal dashed line corresponds to $\Tilde{\lambda}_{min}$, where the wavenumber in the eigenvalue \eqref{lamt} is comparable to the discreteness scale.}
    \label{spectracomp}
\end{figure}

Similarly in the causal set, we see discrete analogues of these eigenfunctions and eigenvalues, up to wavenumbers comparable with the discreteness scale. Figure \ref{spectracomp} shows a comparison of the causal set and continuum eigenvalues of $i\Delta$ for a $2000$ element causal set.\footnote{The eigenvalues in the continuum have dimensions of area while in the causal set they are dimensionless. For the comparison, we have rescaled the causal set eigenvalues using the appropriate density factor.} As evident in the figure, the larger eigenvalues in the causal set and continuum agree well with one another, but at some point, which corresponds to wavenumbers or wavelengths around the discreteness scale (horizontal dashed line in the figure), the causal set eigenvalues start to deviate from a powerlaw. The smaller causal set eigenvalues, appearing as the trend curves down, no longer have continuum-like counterparts. In Section \ref{fluct} we will further discuss the nature of the causal set eigenfunctions corresponding to these small eigenvalues.  

It has been shown \cite{8,20} that the inclusion of the contribution of these small eigenvalues and their eigenfunctions, which do not have continuum-like counterparts, generically leads to spacetime volume laws rather than the conventional spatial area laws for the entanglement entropy. Since a main motivation of the study of entanglement entropy is to understand black hole entropy, it is of great interest to understand if and how the conventional  area laws can be recovered. It has been shown that the area law is recovered after the exclusion of these components from the calculations, using a truncation scheme, which we discuss next. 
\subsection{A Covariant Truncation Scheme}
In a causal set\footnote{For a review of causal set theory, see \cite{2,3,15,23,24}.} with density $\rho=N/V$, where $N$ is the mean number of elements sprinkled into a spacetime  volume $V$, the minimum distance that is meaningful is of the order of the discreteness scale, $\frac{1}{\sqrt{\rho}}$ in $1+1$D. All structures below this scale do not have any physical reality in the causal set. In the present context, knowing what we know about the spectrum of $i\Delta$ and its relation to a wavenumber, we can argue for a maximum meaningful wavenumber (or minimum wavelength) and therefore minimum meaningful eigenvalue in the causal set. 

If, as mentioned already, we recognise that the minimum possible wavelength on the causal set, $\Lambda_{min}$, is given by the discreteness scale
\begin{equation}
    \Lambda_{min}=\frac{1}{\sqrt{\rho}},
\end{equation}
we find that the maximum wavenumber supported on the causal set is
\begin{equation}
    k_{max}=\frac{2\pi}{\Lambda_{min}}=2\pi\sqrt{\rho}.
\end{equation}

Substituting this into the expression for the eigenvalues \eqref{lamt}, we find that the minimum magnitude of an eigenvalue that is meaningful in the causal set is
\begin{equation}
    |\Tilde{\lambda}_{min}^{cont}|=\frac{L}{k_{max}}=\frac{L}{2\pi\sqrt{\rho}},
\end{equation}
where the superscript $cont$ is a reminder that we are expressing the eigenvalues in their dimensionful form, as they appear in the continuum. The eigenvalues in the continuum have dimensions of length squared, while those in the causal set are dimensionless. The two are related to one another through multiplication by a factor of $\rho$. Therefore, in the causal set, we should aim to retain a minimum (in magnitude) eigenvalue 
\begin{equation}
    |\Tilde{\lambda}_{min}^{cs}|= |\Tilde{\lambda}_{min}^{cont}|\times \rho=\frac{\sqrt{N}}{4\pi}.
    \label{mineig}
\end{equation}
The eigenvalues with magnitudes smaller than $\widetilde{\lambda}^{cs}_{min}$ correspond to features below the discreteness scale and should be excluded \cite{20}. This argument assumes that the eigenvalues beyond the bend in Figure \ref{spectracomp} can be continued to be interpreted approximately as $L/k$, even though this is not guaranteed to be the case. In Section \ref{fluct} we will provide further evidence that these contributions have features below the discreteness scale.

Hence, when carrying out the entropy calculation using the formalism of Section \ref{EE}, the spectrum of $i\Delta$ needs to be ``truncated'' such that the magnitude of its minimum eigenvalue is \eqref{mineig}. This truncation is performed at two stages of the calculation. The first is when $i\Delta$ and $W_{SJ}$ are initially defined on the full causal set. The minimum eigenvalues that enter their definition must be consistent with $\eqref{mineig}$. The second truncation is performed after the restriction of $i\Delta$ and $W_{SJ}$ to the subregion (the red subdiamond in Figure \ref{sprinkling2}). The $N$ in \eqref{mineig} corresponds to the number of elements in the region where the truncation is taking place. Hence $N$ would be different in each of the two truncations: in the first truncation it would be the number of elements in the larger diamond $N_L$ and in the second truncation it would be the number of elements in the smaller subdiamond $N_\ell$. Due to the double application of the truncation, this  scheme is sometimes referred to as a ``double truncation''. For more details on the double truncation, see \cite{8,20}. 

As this truncation concerns the smallest scales in the theory, near the discreteness scale, the same truncation can be implemented in the massive scalar field theory provided that $m^2\ll\rho$. We will discuss this further in Section \ref{mive}. It is an open question how to motivate a minimum eigenvalue of the spectrum of $i\Delta$ in general spacetimes, in the absence of analytic results telling us how to relate it to something like a wavenumber (such as in \eqref{lamt}). Our work below in the massive theory, where such analytic results are not available, gives some insight on potential generalisations.

\subsection{Entropy Results}
\label{ent_res}
\begin{figure}[t]
    \centering
    \includegraphics[width=0.55\textwidth]{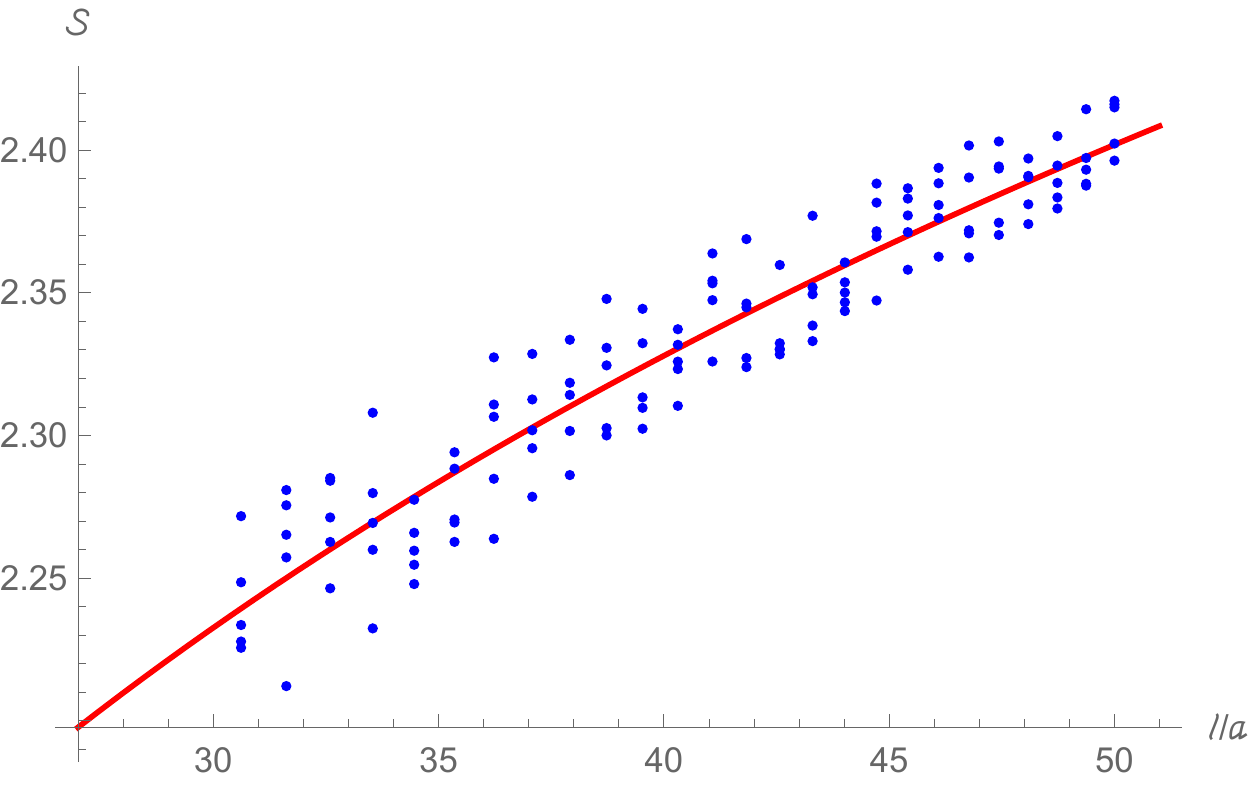}
    \caption{Massless scalar field entanglement entropy versus the half side length $\ell$ of the subdiamond in units of the discreteness length $a$. Sprinklings from $15000$ to $40000$ elements were considered and $\ell/L=1/2$. The data fits $S=0.332 \log{(\ell/a)}+1.104$, shown in red.}
    \label{massless_scaling}
\end{figure}

\begin{figure}[h!]
        \centering
        \begin{subfigure}[b]{0.475\textwidth}
            \centering
            \includegraphics[width=\textwidth]{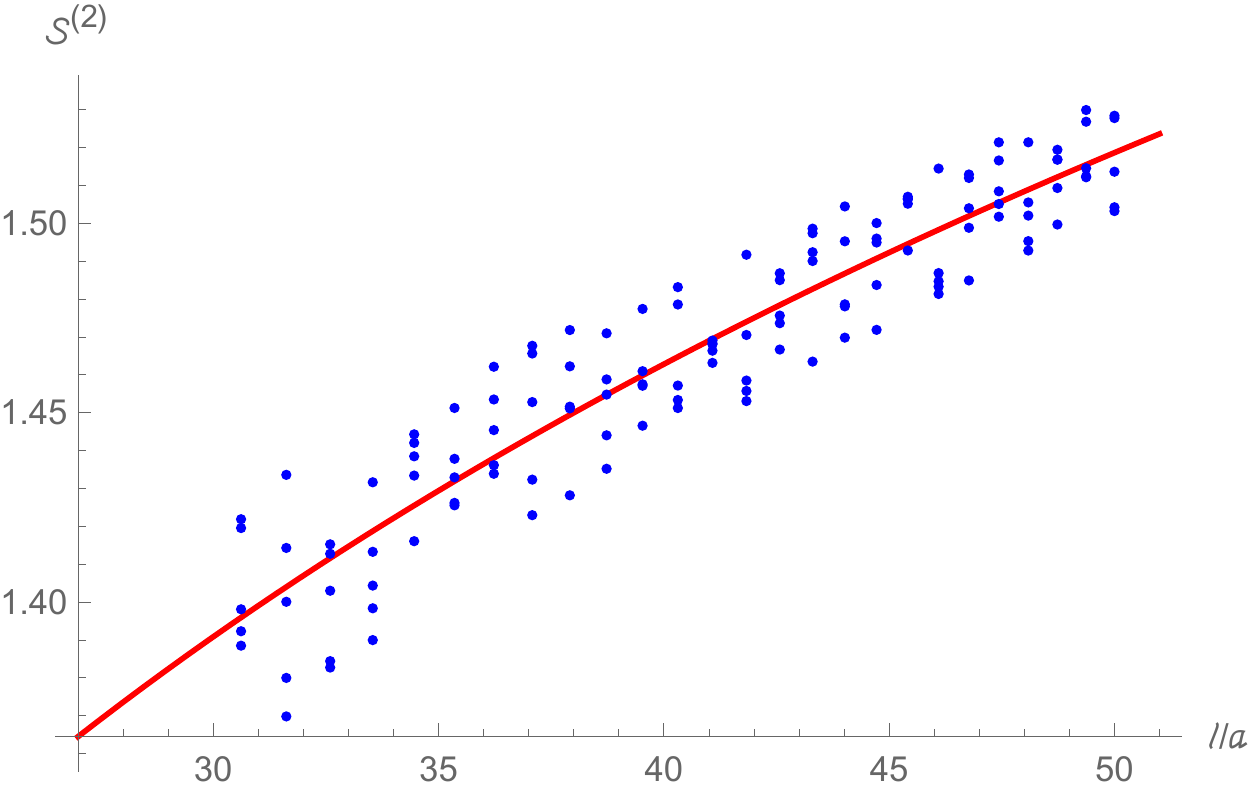}
            \caption[Network2]%
            {{Rényi entropy of order 2 with a best fit of $S^{(2)}=0.250\log{(\ell/a)}+0.540$.}}    
            \label{renyi2}
        \end{subfigure}
        \hfill
        \begin{subfigure}[b]{0.475\textwidth}  
            \centering 
            \includegraphics[width=\textwidth]{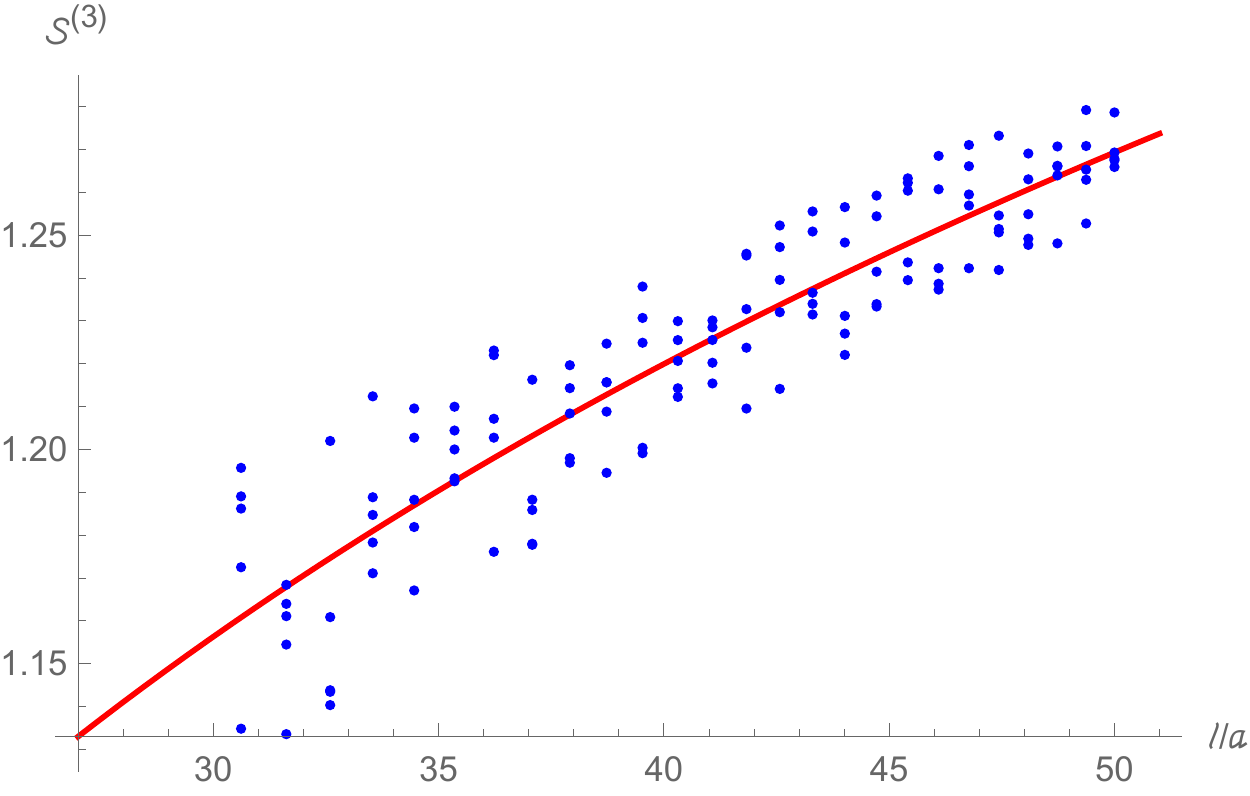}
            \caption[]%
            {{Rényi entropy of order 3 with a best fit of $S^{(3)}=0.223\log{(\ell/a)}+0.404$.}}    
            \label{renyi3}
        \end{subfigure}
        \vskip\baselineskip
        \begin{subfigure}[b]{0.475\textwidth}   
            \centering 
            \includegraphics[width=\textwidth]{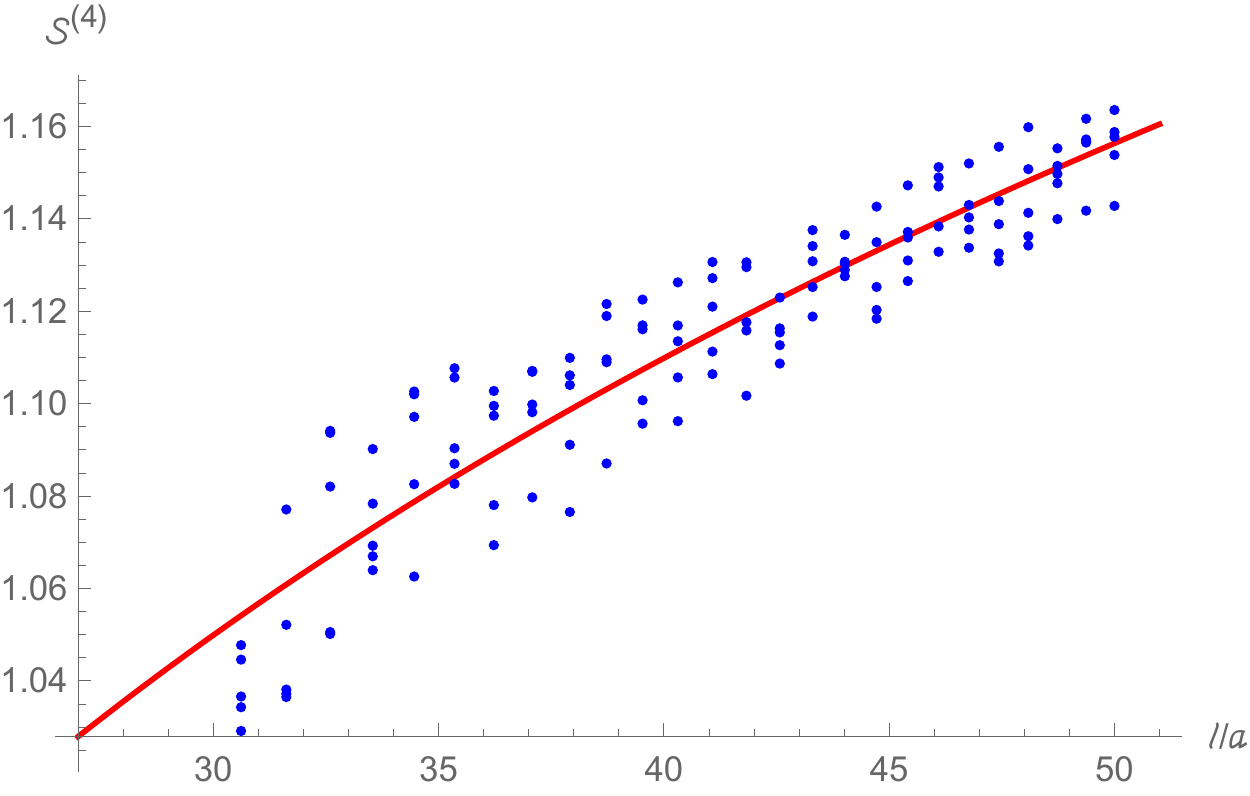}
            \caption[]%
            {{Rényi entropy of order 4 with a best fit of $S^{(4)}=0.208\log{(\ell/a)}+0.341$.}}    
            \label{renyi4}
        \end{subfigure}
        \hfill
        \begin{subfigure}[b]{0.475\textwidth}   
            \centering 
            \includegraphics[width=\textwidth]{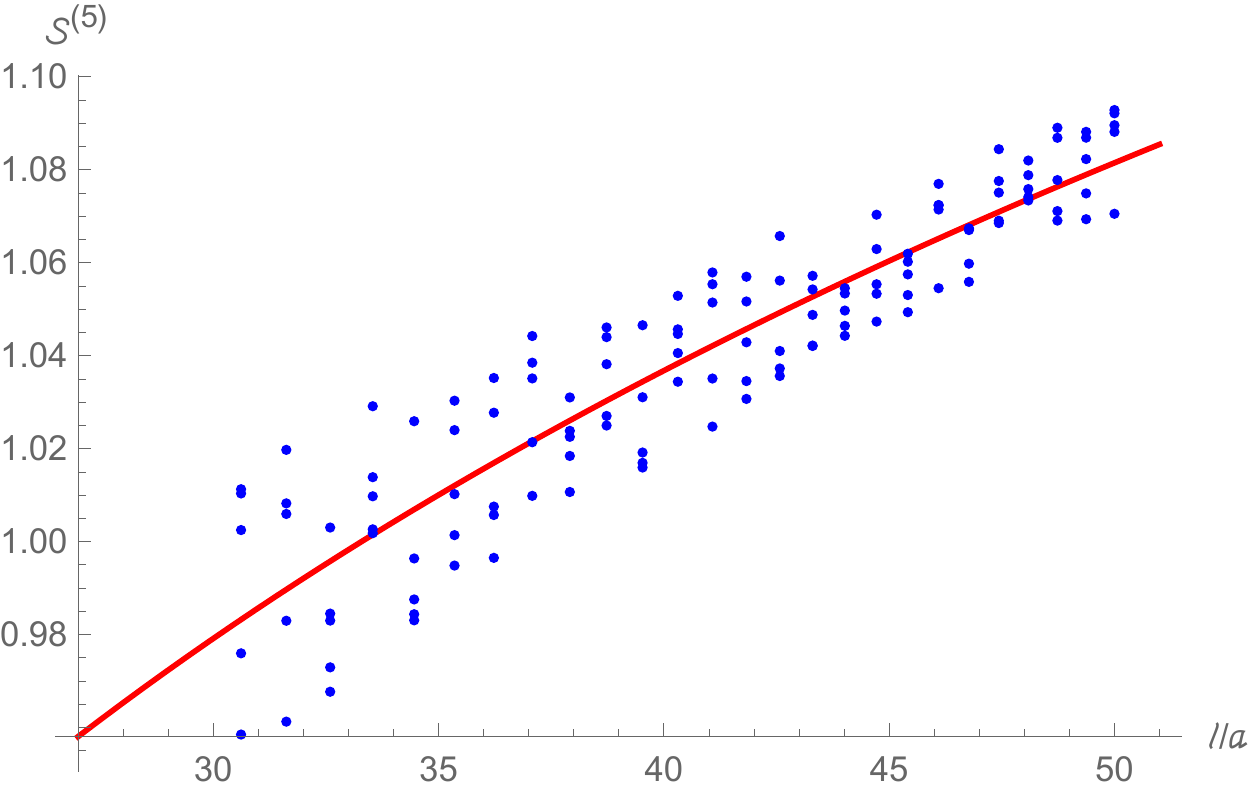}
            \caption[]%
            {{Rényi entropy of order 5 with a best fit of $S^{(5)}=0.200\log{(\ell/a)}+0.298$.}}    
            \label{renyi5}
        \end{subfigure}
        \caption{Rényi entropies of order 2, 3, 4, and 5 for a massless scalar field on a causal set. Sprinklings from $15000$ to $40000$ elements were considered and $\ell/L=1/2$. The best fit curves are shown in red.} 
        \label{renyi}
    \end{figure}

The causal set setup of our calculations is Figure \ref{sprinkling2}. The UV cutoff $a$ that we will study the entanglement entropy scalings with respect to is the discreteness scale $a\equiv 1/\sqrt{\rho}$. We will vary $a$ by holding fixed the volumes of the causal diamonds ($\ell/L=1/2$ and $L=1/ 2$) and varying the number of elements sprinkled into them. We consider 130 sprinklings from $15000$ to $40000$ elements in increments of $1000$ with 5 sprinklings per increment. The results for the entanglement entropy \eqref{see} and R\'enyi entropies \eqref{ftr} of order $2$ to $5$ are shown in Figures \ref{massless_scaling} and \ref{renyi}.

Since the causal diamonds in Figure \ref{sprinkling2} are the domains of dependence of their spatial diameters, our results for the entanglement entropy and R\'enyi entropy can be compared to \eqref{area} and \eqref{cardy_renyi} respectively. In all our calculations we used \eqref{see} and \eqref{ftr} and applied the double truncation scheme described in the previous subsection. 

The results for the entanglement entropy versus $\ell/a$ are shown in  Figure \ref{massless_scaling}. In agreement with the results of \cite{20} and \eqref{area}, a logarithmic scaling with a coefficient consistent with the expected $1/3$ is obtained. More precisely, the scaling coefficient is $0.332 \pm 0.009$ and the non-universal constant is $1.104 \pm 0.037$, where the uncertainties were calculated by taking the square root of the diagonal elements of the covariance matrix of the fit.

The results for the R\'enyi entanglement entropies $S^\alpha$ of orders $\alpha=2,3,4,5$ are shown in Figures \ref{renyi2}, \ref{renyi3}, \ref{renyi4},  and \ref{renyi5} versus $\ell/a$. Once again, logarithmic scalings are obtained, with coefficients $0.250 \pm 0.008$, $0.223 \pm 0.009$, $0.208 \pm 0.007$, and $0.200 \pm 0.008$, in good agreement with the expected coefficients of $\frac{1}{4}=0.25$, $\frac{2}{9}=0.222$, $\frac{5}{24}=0.208$, and $\frac{1}{5}=0.2$ respectively from \eqref{cardy_renyi}. The non-universal constants are $0.540 \pm 0.030$, $0.404 \pm 0.032$, $0.341 \pm 0.027$, $0.298 \pm 0.028$.\footnote{In future work it would be interesting  to study these non-universal constants in greater detail.}

\clearpage
    \clearpage

\section{Massive Scalar Field Theory}
\label{mive}
The retarded Green function for a massive scalar field in $1+1$D Minkowski spacetime is 
\begin{equation}\label{eq:2.11}
G_{R,m}^{(2)}(x,\,x') = \frac{1}{2}\theta(t-t')\theta(-\tau^2)J_0(m\sqrt{-\tau^2}),
\end{equation}
where $J_0$ is the zeroth order Bessel function of the first kind \cite{16}. Notice that there is a non-trivial dependence on the proper time, compared to the massless expression \eqref{eq:2.10}. The causal set retarded Green function for the massive theory is  \cite{1}
\begin{equation}\label{eq:2.12}
G_{R,xx'} = \frac{1}{2}C_{x'x}\Bigg(I +\frac{m^2}{2\rho}C_{x'x} \Bigg)^{-1},
\end{equation}
where $m^2\ll\rho$, and $I$ is the identity matrix.

The Pauli-Jordan function in the continuum theory is
\begin{equation}
\label{pjm}
    \Delta(u,v;u',v')=-\frac{1}{2}(\theta(u-u')+\theta(v-v')-1)J_0\Big(m\sqrt{2(u-u')(v-v')}\Big).
\end{equation}

\begin{figure}[t!]
\centering
\begin{subfigure}{.5\textwidth}
  \centering
  \includegraphics[width=\linewidth]{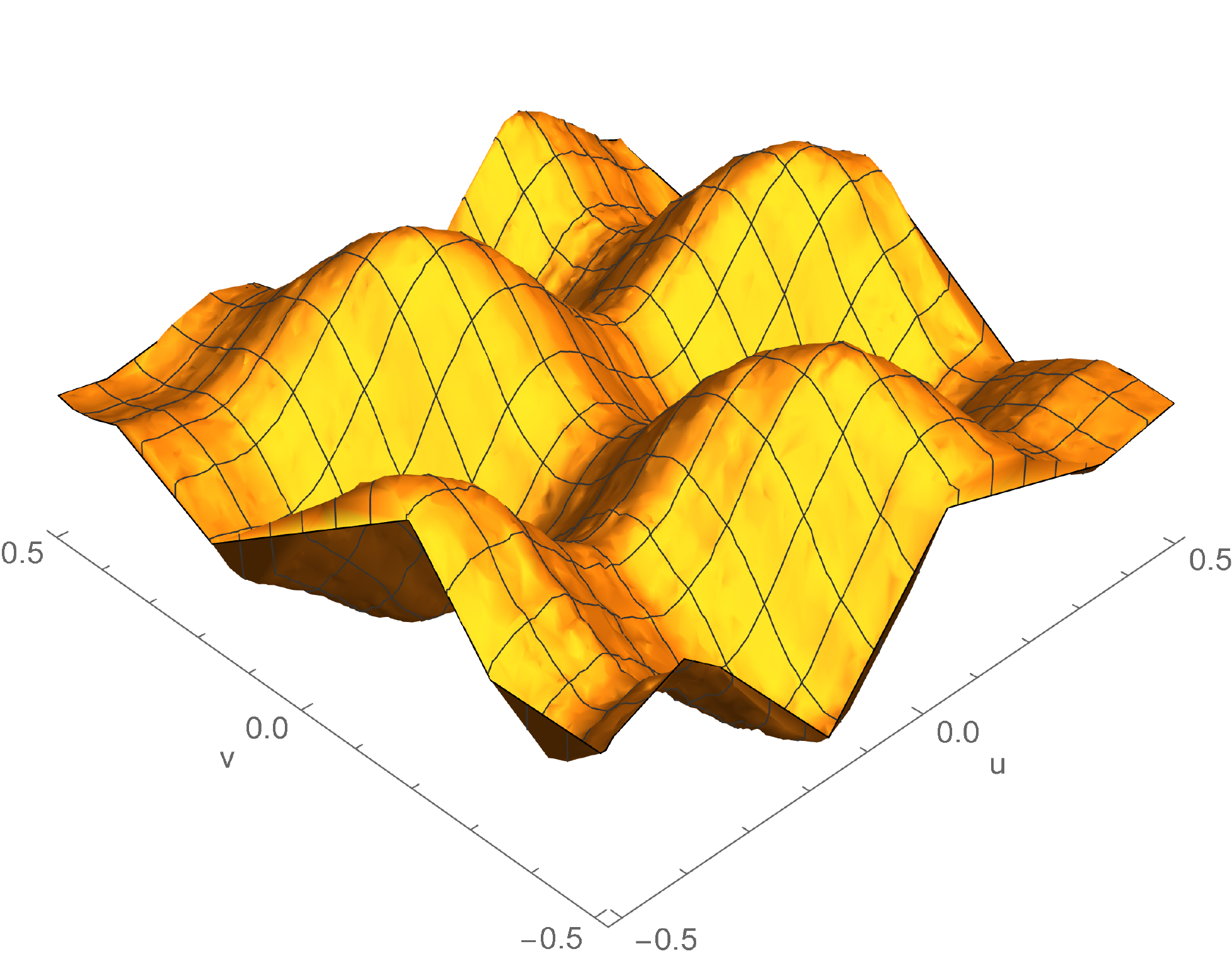}
  \caption{}
  \label{fig:sub1}
\end{subfigure}%
\begin{subfigure}{.5\textwidth}
  \centering
  \includegraphics[width=0.95\linewidth]{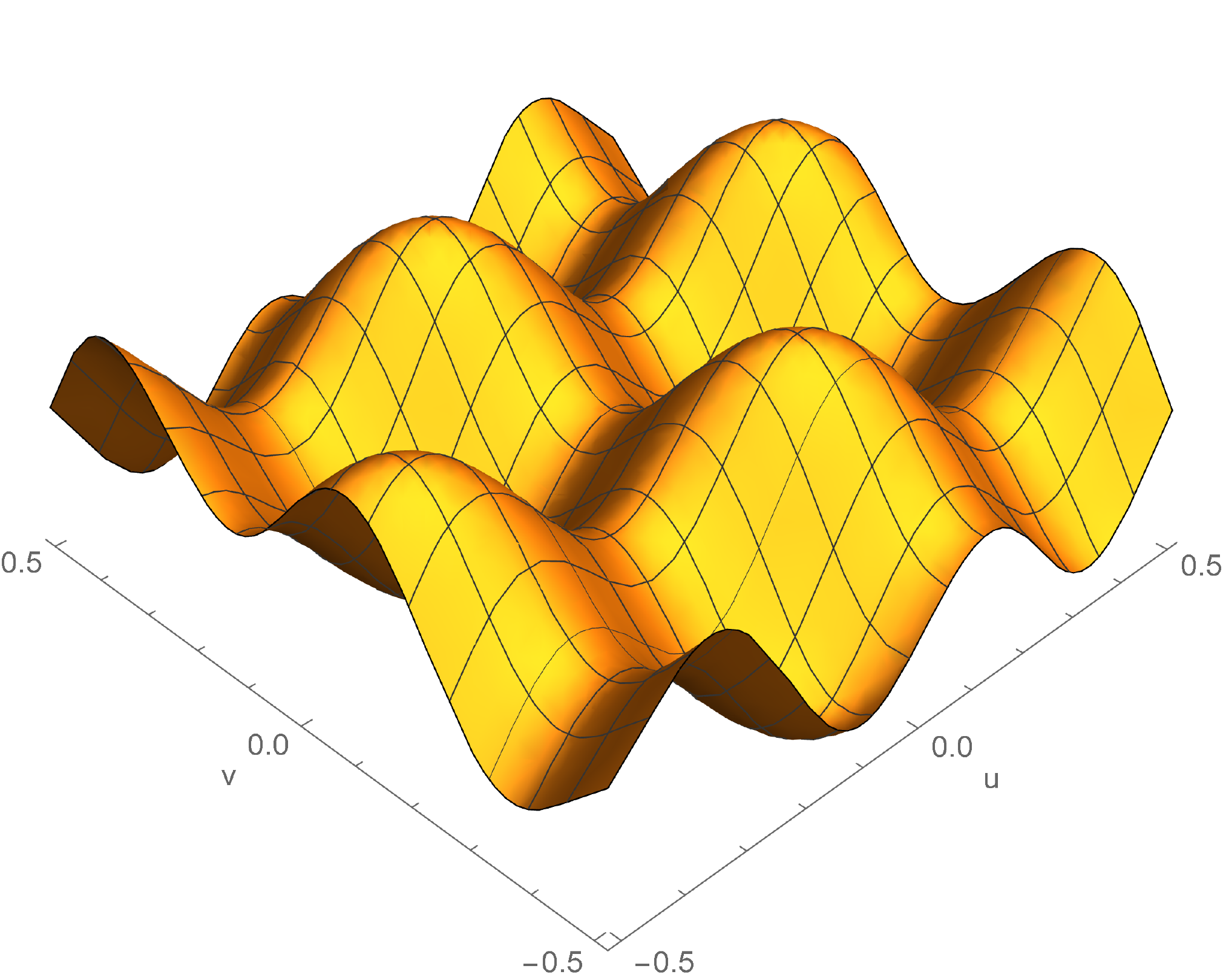}
  \caption{}
  \label{fig:sub2}
\end{subfigure}
\caption{Left: a plot of the interpolation of the real part of an eigenfunction of $i\Delta$ for a massive scalar field with $m=10$ on a causal set sprinkling of 10000 elements with $2L=1$, against $u$ and $v$. Right: a plot of $\text{Re}\{f_{k,m}\}$ with a guess value of $k=7$ against $u$ and $v$.}
\label{eigcomp}
\end{figure}

Due to presence of the Bessel function in \eqref{pjm}, the integral eigenvalue problem  \eqref{eigvalprob} in a causal diamond is considerably more complicated in the massive theory. Instead of trying to solve for the eigenfunctions directly, we can extrapolate \eqref{eigmassless} to the massive case by introducing $\omega^2=|k|^2+m^2$. This analogy gives
\begin{flalign}\label{eigmassivewrong}
\begin{split}
&f_{k,m}(u,\,v):=e^{-i\frac{(w+k)u}{\sqrt{2}}}e^{-i\frac{(w-k)v}{\sqrt{2}}}-e^{-i\frac{(w-k)u}{\sqrt{2}}}e^{-i\frac{(w+k)v}{\sqrt{2}}}\\
&g_{k,m}(u,\,v):=e^{-i\frac{(w+k)u}{\sqrt{2}}}e^{-i\frac{(w-k)v}{\sqrt{2}}}+e^{-i\frac{(w-k)u}{\sqrt{2}}}e^{-i\frac{(w+k)v}{\sqrt{2}}}-2\text{cos}(\sqrt{2}kL).
\end{split}
\end{flalign} 
These eigenfunctions can then be empirically compared to those obtained directly from the causal set, by substituting in guess values of $k$, as shown in Figure \ref{eigcomp} for a massive scalar field with $m=10$ on a causal set sprinkling of 10000 elements with $2L=1$. The eigenfunction shown corresponds to the first antisymmetric eigenfunction and was thus compared to $f_{k,m}$ with a guess value of $k=7$.
\begin{figure}[h]
    \centering
    \includegraphics[width=\linewidth]{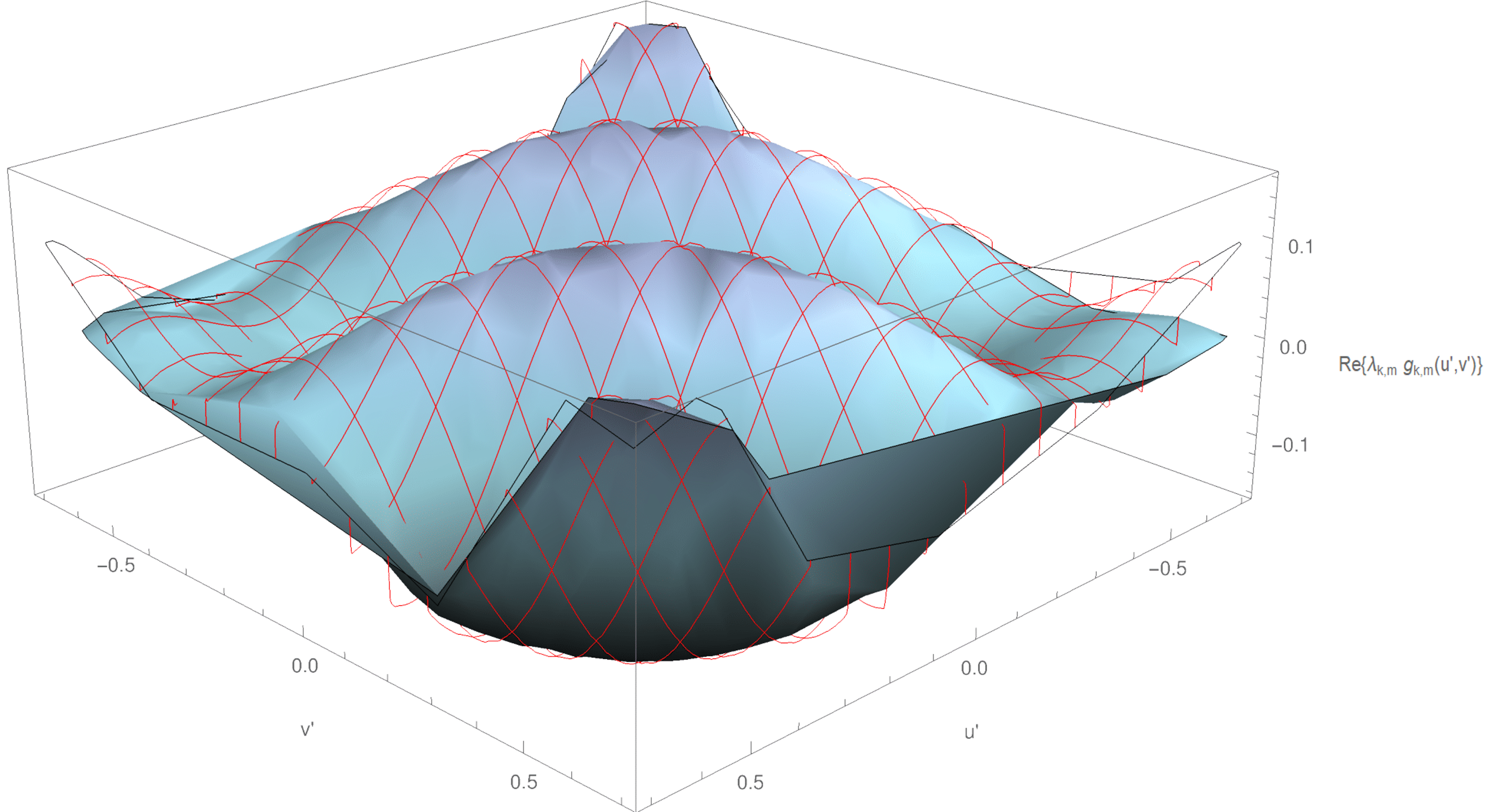}
    \caption{Plot of $\text{Re}\{i\Delta(u-u',v-v')g_{k,m}(u,v)\}$ in blue and $\text{Re}\{\Tilde{\lambda}_{k,m} \, g_{k,m}(u',v')\}$ in red mesh against $u'$ and $v'$ for $k=3$ and $m=10$.}
    \label{eigcomp3d}
\end{figure}

However, in contrast to the massless case, it was found through numerical integration that the eigenvalue equation \eqref{eigvalprob} is no longer fulfilled with the $cos$ term included in the $g$ family of eigenfunctions. One should also notice that \eqref{eigmassivewrong} is not a solution of the massive Klein-Gordon equation due to the presence of the $cos$ term. Hence, \eqref{eigmassivewrong} should be modified to
\begin{flalign}\label{eigmassiveright}
\begin{split}
&f_{k,m}(u,\,v):=e^{-i\frac{(w+k)u}{\sqrt{2}}}e^{-i\frac{(w-k)v}{\sqrt{2}}}-e^{-i\frac{(w-k)u}{\sqrt{2}}}e^{-i\frac{(w+k)v}{\sqrt{2}}}\\
&g_{k,m}(u,\,v):=e^{-i\frac{(w+k)u}{\sqrt{2}}}e^{-i\frac{(w-k)v}{\sqrt{2}}}+e^{-i\frac{(w-k)u}{\sqrt{2}}}e^{-i\frac{(w+k)v}{\sqrt{2}}}.
\end{split}
\end{flalign}
We also find that for \say{the right values of $k$}, the left and right hand sides of the eigenvalue equation \eqref{eigvalprob} are in almost complete agreement if we keep away from the left and right corners of the diamond, as shown in Figure \ref{eigcomp3d}.\footnote{This is also consistent with the result that in the infinite volume spacetime, the SJ state is the Minkowski vacuum and the positive (negative) eigenvalue SJ modes are linear combinations of positive (negative) frequency plane wave solutions to the Klein Gordon equation \cite{30}.} In this case, the eigenvalues, $\Tilde{\lambda}_{k,m}$ are given by 
\begin{equation}\label{lambdamassive}
\Tilde{\lambda}_{k,m} \simeq \pm \frac{\sqrt{2}L}{\omega + |k|},   
\end{equation}
where the $\sqrt{2}$ factor can be understood simply by setting $m=0$ in \eqref{eigmassiveright}. While we were unable to find exact and closed form expressions for the eigendecomposition of $i\Delta$ in the massive theory, our approximate results are encouraging. The form of the eigenvalues $\Tilde{\lambda}_{k,m}$, \eqref{lambdamassive}, tells us that as long as $m^2\ll \rho$, the massive theory's eigenvalues  approach the massless theory's eigenvalues in the UV limit (as the eigenvalues become smaller). This is exactly the regime we need to understand in order to apply a meaningful truncation scheme for the entanglement entropy. Therefore, since the spectrum \eqref{lambdamassive} approaches the massless one \eqref{lamt} in the UV, in the massive theory we can expect to retain the same minimum magnitude eigenvalue as in the massless theory.

\begin{figure}[h]
    \centering
    \includegraphics[width=0.7\linewidth]{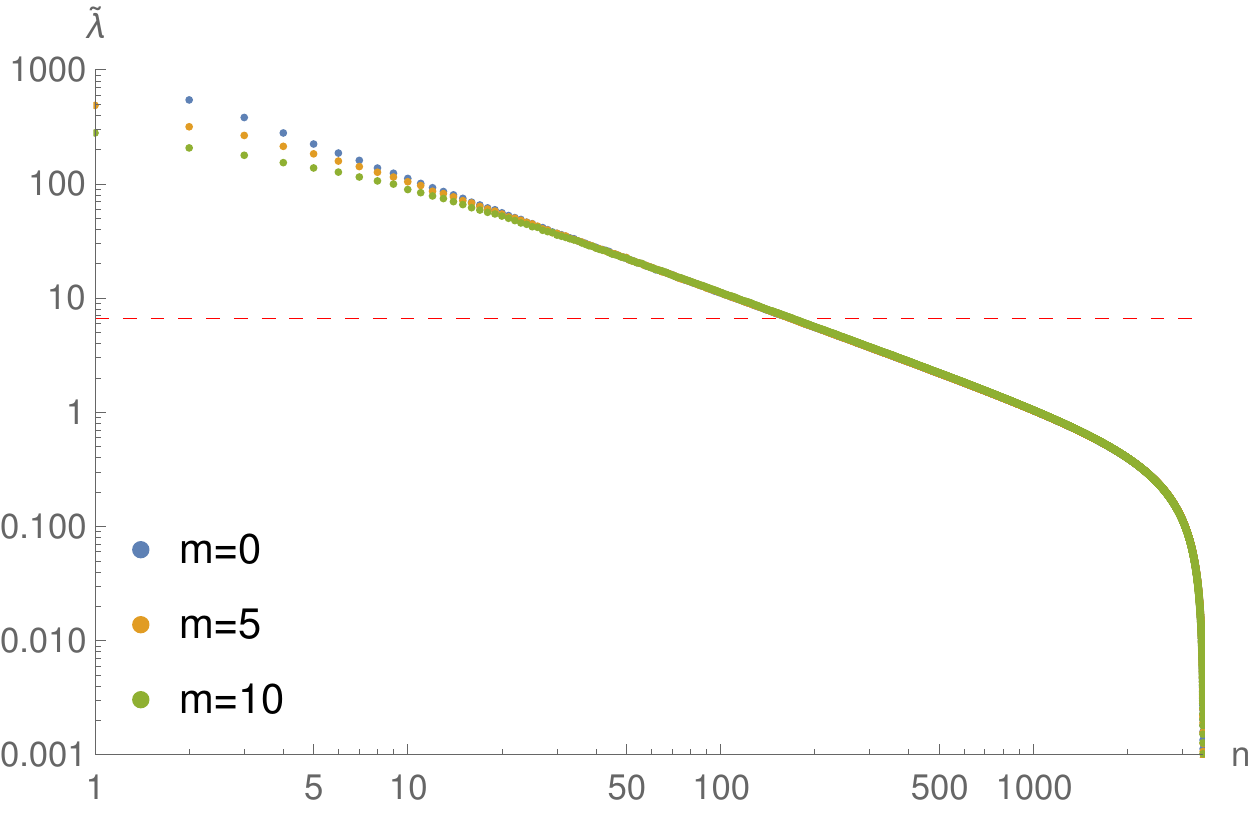}
    \caption{Log-Log plot of each $n^{th}$ largest $\Tilde{\lambda}$ against $n$, for $L=1/2$ and a sprinkling density of $\rho=7000$ for $m=0$ in blue, $m=5$ in orange, and $m=10$ in green. The red dashed horizontal line corresponds to $\Tilde{\lambda}_{min}$.}
    \label{7000massive}
\end{figure}

This is also illustrated in Figure \ref{7000massive}, where the spectrum of $i\Delta$ for two massive theories ($m=5$ and $m=10$) are shown together with spectrum in the massless theory. The eigenvalues of $i\Delta$ are only significantly different for large $\Tilde{\lambda}$ and are essentially the same when  $\Tilde{\lambda} \sim \Tilde{\lambda}_{min}$.

For small values of $k$ (or large $\Tilde{\lambda}_{k,m}$), the mass plays a significant role in determining the magnitude of $\Tilde{\lambda}_{k,m}$. However, for larger values of $k$, unless the mass is itself of the same order as $\sqrt{\rho}$ (which would not be a well-defined theory in the causal set), $\omega=\sqrt{|k|^2+m^2}\sim k$. Thus, as $\Tilde{\lambda}_{min}$ corresponds to a UV limit, for $m \ll \sqrt{\rho}$, we have that
\begin{equation}\label{lambdamassive3}
    |\Tilde{\lambda}_{min,m}| \sim |\Tilde{\lambda}_{min}|=\frac{\sqrt{N}}{4\pi},
\end{equation}
where $\Tilde{\lambda}_{min,m}$ corresponds to the minimum eigenvalue to retain in the causal set truncations in the massive case. This is what we use in our calculations below.

\subsection{Entropy Results}
\begin{figure}[b]
    \centering
    \includegraphics[width=0.7\linewidth]{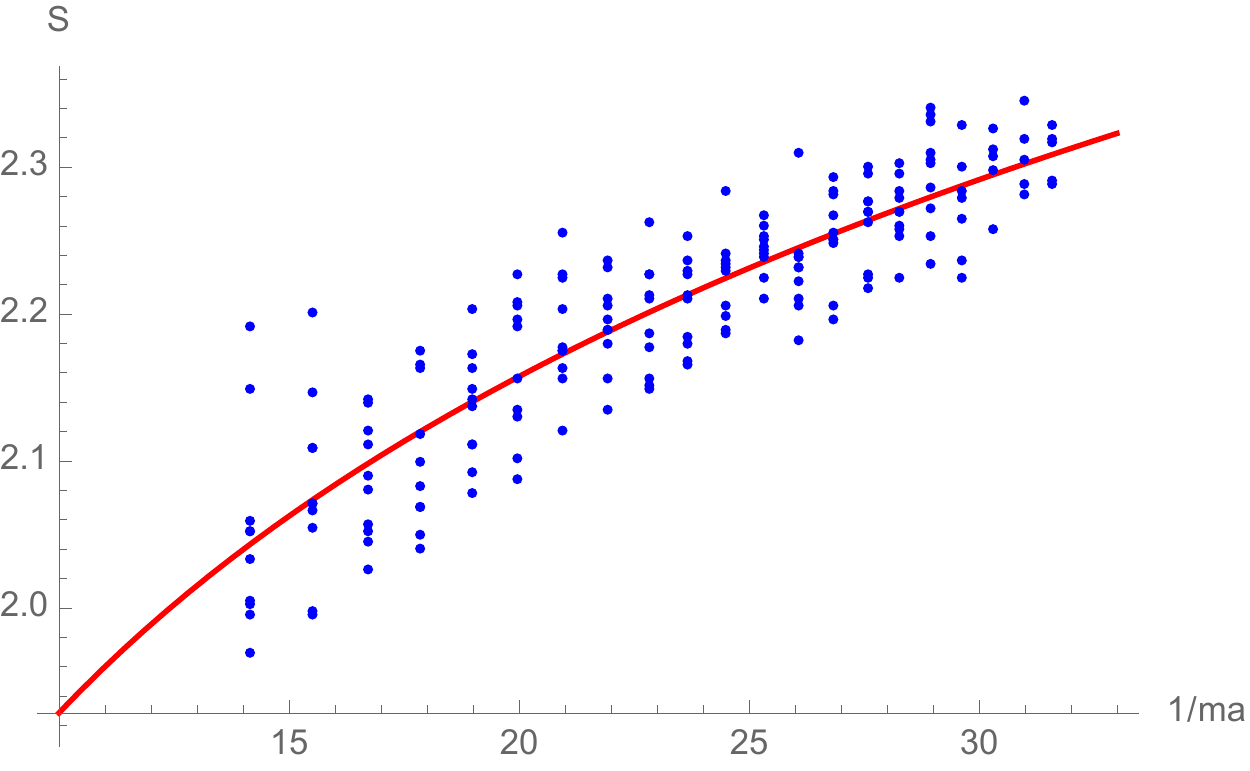}
    \caption{Fit of $S= 0.33 \log{(1/ma)}+ 1.17$ for a massive scalar field on causal set sprinklings of up to 25000 elements with $\ell/L=1/2$, and $m=5$.}
    \label{massive_final_a}
\end{figure}

The setup of our calculations in the massive theory is the same as in Section \ref{ent_res}. As motivated in the previous subsection, we use the same minimum eigenvalue truncation scheme as in the massless theory. First, we hold the mass fixed ($m=5$) and vary $a$. Once the truncations are applied, the scaling with respect to the UV cutoff $a=1/\sqrt{\rho}$ follows $S=0.33 \log{(1/ma)}+ 1.17$, in agreement with the expected form \eqref{cardy_massive}. These results are shown in Figure \ref{massive_final_a} along with the best fit log scaling with a coefficient of $0.33 \pm 0.01$ and non-universal constant $1.17 \pm 0.04$.

\begin{figure}[t]
    \centering
    \includegraphics[width=0.7\linewidth]{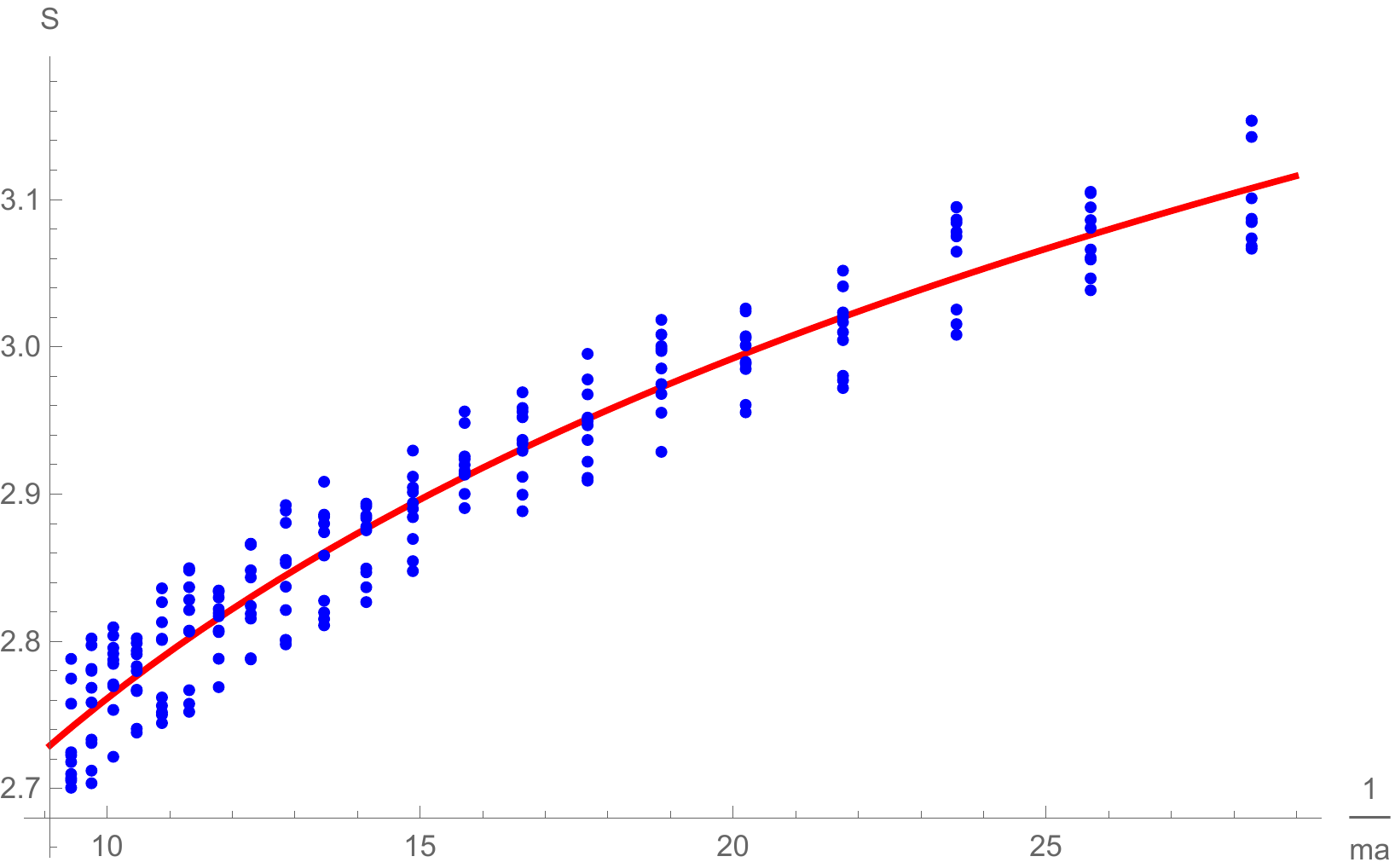}
    \caption{Fit of $S=0.333 \log{(1/ma)}+1.993$ for a massive scalar field on causal set sprinklings of 20000 elements with $\ell/L=1/2$,  and masses ranging from $5$ to $15$.}
    \label{massive_final}
\end{figure}

We also studied the entanglement entropy scaling with respect to the mass, in the range $5\leq m\leq 15$ in a diamond with side length $2L=1$ and $\rho=20000$. The results are shown in Figure \ref{massive_final}, along with the best fit log scaling $S=0.333 \log{(1/ma)}+1.993$. The coefficient of the log fit, $0.333 \pm 0.006$, is in good agreement with the expected result \eqref{cardy_massive}. The non-universal constant is $1.993 \pm 0.017$.

We have thus extended some of the main results of \cite{20,8} to the massive scalar field theory. Knowledge of the form of the eigenvalues in the massive case was vital in obtaining a physical understanding of the spectrum of $i\Delta$, from which a meaningful truncation scheme was defined. As shown above, we obtained the expected ($1+1$D log) ``area law'' scalings   for the massive scalar field theory. The universality of the spectrum of $i\Delta$ in the UV regime, where truncations need to be implemented, may hold true more generally. This could make extensions to more general theories and spacetimes possible as long as the UV limit of the massless theory in D-dimensional flat spacetime is well understood. We discuss this further in Section \ref{summary}.

\section{Fluctuations}
\label{fluct}
So far we have focused on the eigenvalues that we wish to keep in the truncation (the left of the bend in the spectrum of Figures \ref{spectracomp} and \ref{7000massive}), and their corresponding eigenfunctions which are plane wave-like. In this Section we turn our attention to the discarded eigenvalues and eigenfunctions. 

From the spectra in Figures \ref{spectracomp} and \ref{7000massive}, it is already evident that the latter set has a qualitatively different behaviour. The spectrum no longer follows a power law in this regime. This is seen by the trend in the Figures \ref{spectracomp} and \ref{7000massive} curving down sharper and sharper as we move to the right.

\begin{figure}
     \centering
     \begin{subfigure}[h]{0.8\textwidth}
         \centering
         \includegraphics[width=\textwidth]{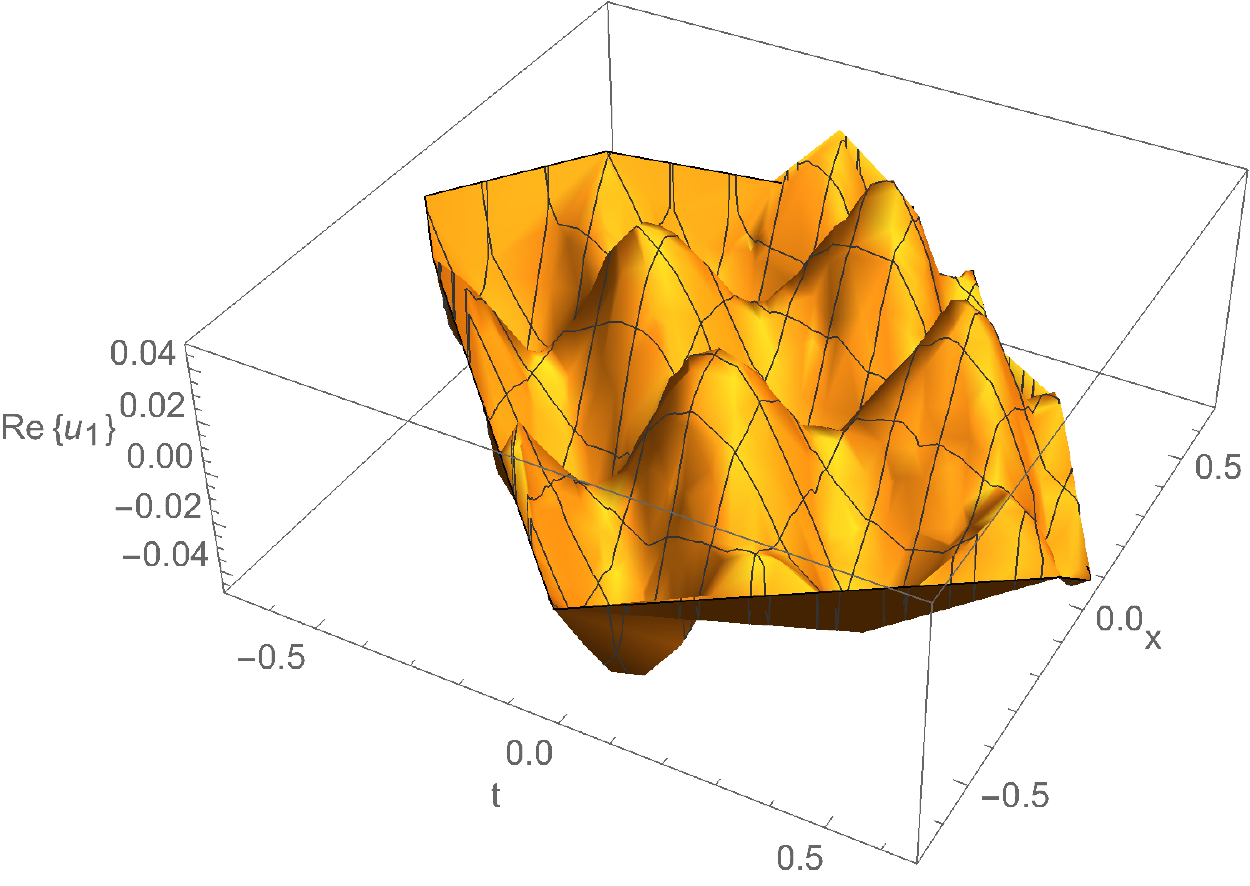}
         \caption{Real part of the eigenvector corresponding to the 10$^{th}$ largest eigenvalue for a causal set of 1000 elements and $L=1/2$.}
         \label{fig:y equals x}
     \end{subfigure}
     \hfill
     \begin{subfigure}[h]{0.8\textwidth}
         \centering
         \includegraphics[width=\textwidth]{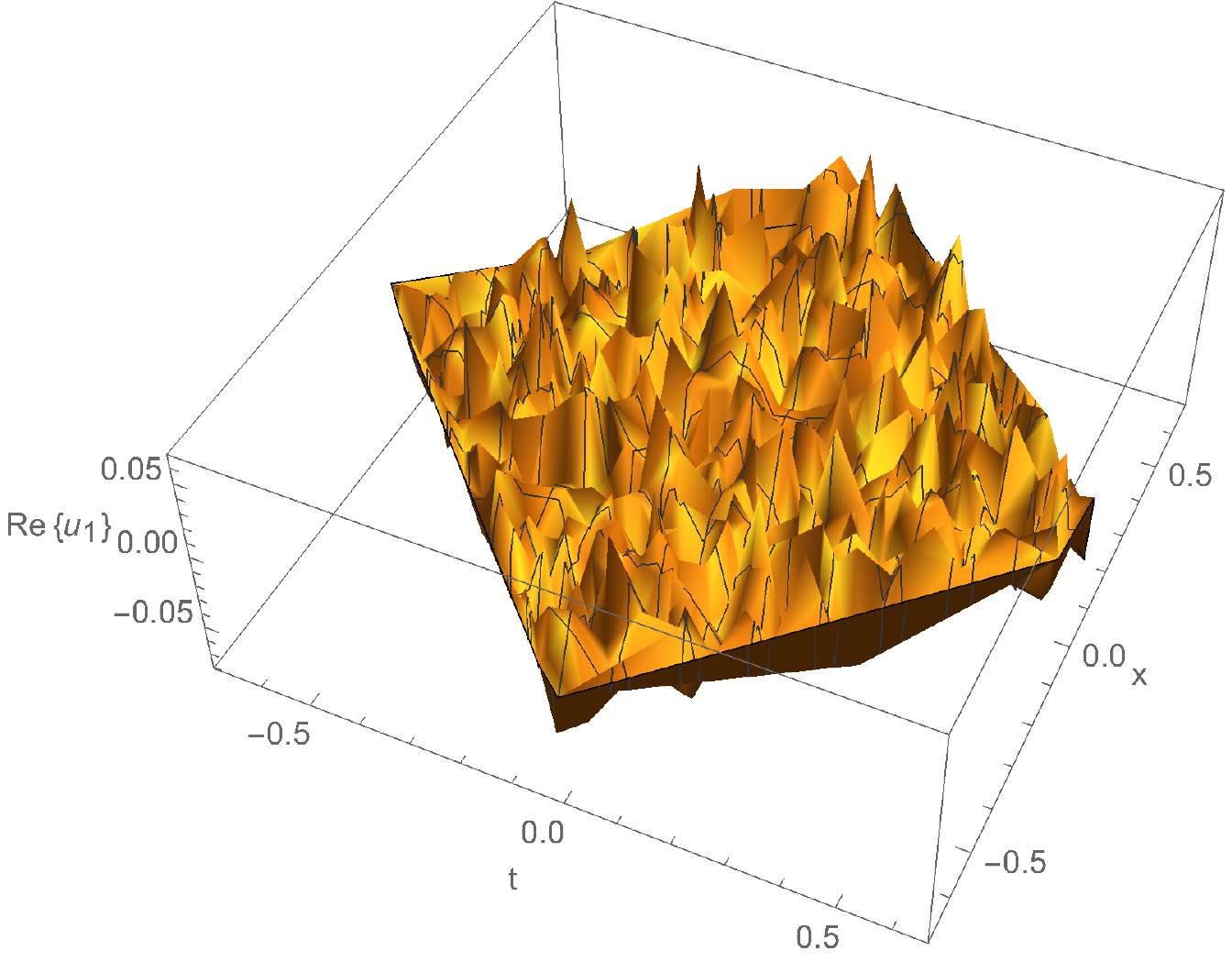}
         \caption{Real part of the eigenvector corresponding to the 425$^{th}$ largest eigenvalue for a causal set of 1000 elements and $L=1/2$.}
         \label{fig:three sin x}
     \end{subfigure}
        \caption{Comparison of the eigenvectors corresponding to a large and a small eigenvalue of $i\Delta$.}
        \label{jagged}
\end{figure}

It is also instructive to compare a typical eigenvector corresponding to a large (within the power law regime) and a small (past the power law regime) eigenvalue in the causal set. These are illustrated in Figure \ref{jagged}. At the top, in Figure \ref{fig:y equals x}, is the real part of the eigenvector corresponding to the $10^{th}$ largest positive eigenvalue in a massless theory on a causal set with $N=1000$ elements and a diamond with $L=1/2$. The eigenvector values have been interpolated between the values they take on causal set elements. As expected from \eqref{eigmassless}, the behaviour is consistent with a linear combination of plane waves, and the oscillations are smooth and above the discreteness scale. At the bottom, in Figure \ref{fig:three sin x}, is the real part of the eigenvector associated with the $425^{th}$ largest positive eigenvalue in the same causal set. This eigenvector has a much more jagged behaviour, with features near the discreteness scale, and does not resemble a linear combination of plane waves with variations above the discretenesss scale. The comparison of the imaginary parts of the eigenvectors follows the same trend. While there is no objective or unique measure of smoothness or jaggedness, the qualitative contrast of smooth oscillations above the discreteness scale versus sharp variations over scales at or below the discreteness scale is a persistent feature of the two branches of eigenvectors in the causal set. 

The jagged and chaotic nature of the small eigenvalue components prompts the question of whether they can be viewed as some kind of fluctuation arising from a particular causal set sprinkling. We investigate this question as follows: We consider a fixed coarse grained (by $10\%$) causal set of $N_c$ elements sprinkled into a diamond with $L=1/2$. We then consider an ensemble of $20$ denser sprinklings with $N$ elements (where $N_c/N=9/10$), including the original $N_c$. Namely, the $N_c$ elements will be shared by all causal sets in the ensemble, while the remaining $N-N_c$ elements will be different. For each of these causal sets, we compute the eigenvectors and eigenvalues of $i\Delta$ for the massless scalar theory. We then use the following  strategy to study whether an eigenvector is a fluctuation occurring in a particular causal set or else an eigenvector with physical reality in every causal set in the ensemble:
\setcounter{footnote}{0}
First, for each sprinkling, we average the $n=10$ largest (in absolute value) entries of each $j^{th}$ eigenvector.\footnote{The eigenvectors are ordered based on their eigenvalues, from largest to smallest in magnitude.} Then, for each $j$, we take the average of this quantity over all the sprinklings in the ensemble. 

Second, we set the $j^{th}$ eigenvector of each sprinkling to the same phase. For example, this can be done by requiring the imaginary part to vanish at the same fixed element (which must be one of the elements in the coarser causal set) in all the sprinklings. Then, the average of each  in-phase $j^{th}$ eigenvector on the $N_c$ elements is taken across all the sprinklings, producing an ``averaged eigenvector''. This averaging is possible because we have fixed the $N_c$ elements across all the sprinklings. Had we used entirely different sprinklings, there would not be a meaningful notion of taking the average at an element(s), because the different elements in the different sprinklings cannot not be identified with one another. Next, we repeat a similar procedure as in the first step: we average the absolute values of the $n=10$ largest entries of the averaged eigenvectors. This boils down the properties of our averaged eigenvector to a single number, which we can compare to the analogous number we obtained in step one (where we didn't first average over $N_c$).

\begin{figure}[h!]
\centering
\includegraphics[scale=0.8]{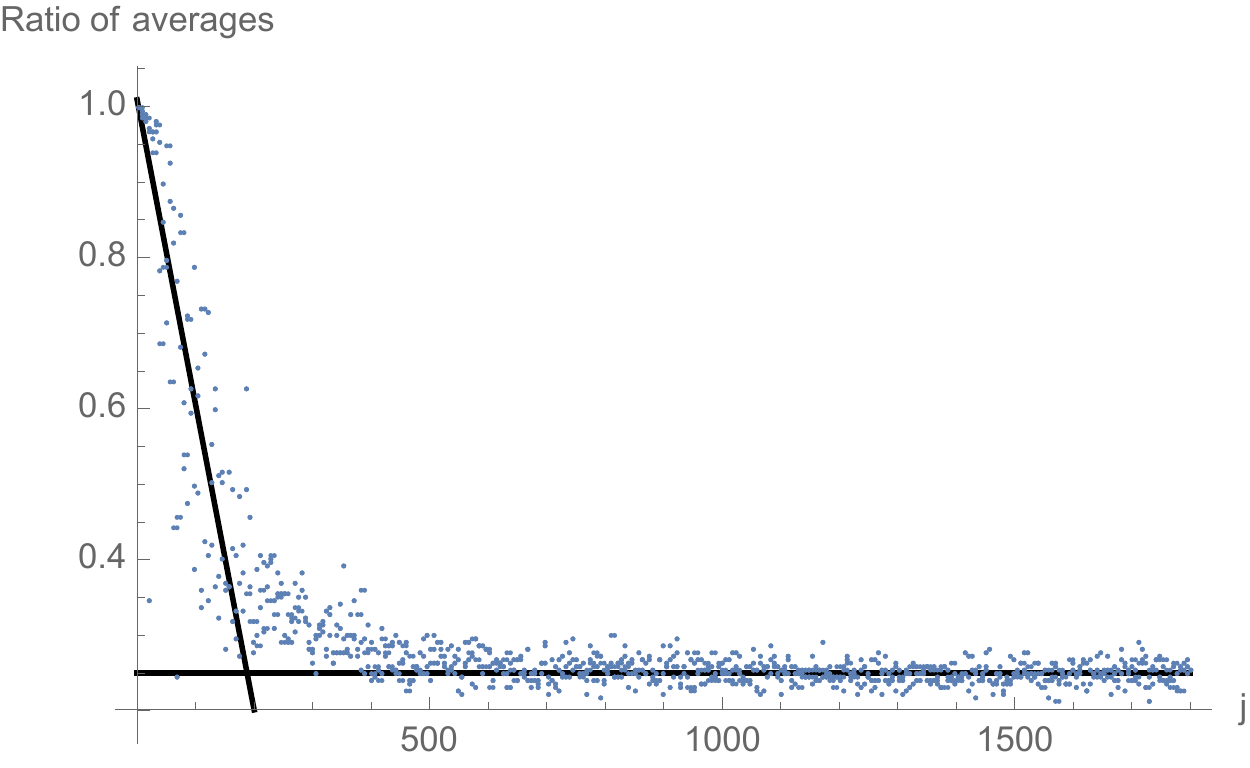} 
\caption{Ratio of averages for causal sets of $N=2000$ elements, along with a constant and a linear fit. $j$ corresponds to the number of an eigenvalue (when ordered from largest to smallest).}
\label{averages}
\end{figure}

Finally, we divide the result obtained in the second step by the result obtained in the first step. An example of the results of this process are shown in Figure \ref{averages}. A ratio close to $1$ indicates that the behaviour of a single eigenvector captures well other  sprinklings' eigenvectors corresponding to the same eigenvalue. On the other hand, lower ratios show that the eigenvectors are more particular to a specific causal set, i.e. they fluctuate and cancellations occur, leading to an averaged vector with a smaller magnitude. The latter could be interpreted as a ``fluctuation'' arising in a particular sprinkling.

As evident in Figure \ref{averages}, the ratio first monotonically decreases before approaching a constant. This means that there is an eigenvalue beyond which the eigenvectors do not become more ``jagged'' or fluctuation-like, according to our prescription.

\subsection{Relation to Truncations}
The trend in Figure \ref{averages} shows a steady decrease, signalling an approach to more fluctuation-like behaviour, followed by a stagnation in the fluctuation-like regime. These two regimes of decrease and stagnation are approximately marked by lines drawn in the figure. An interesting question is, could the transition between these two regimes be related to the transitions in Figures \ref{spectracomp} and \ref{7000massive} from the power law regime to the non-power law regime? Viz, could this transition be another signature of the point at which we must truncate? Let us investigate this.

\begin{figure}[h!]
\centering
\includegraphics[scale=0.8]{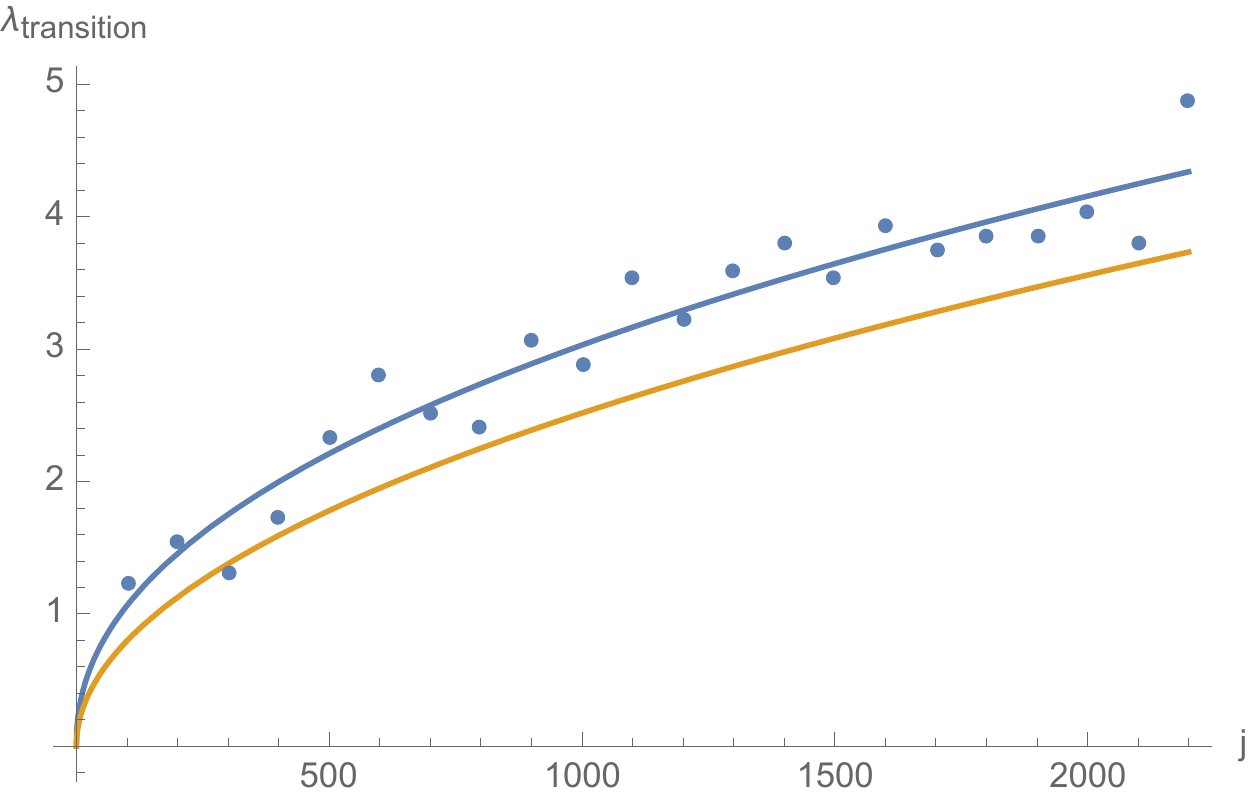}
\caption{$\lambda_{transition}$ at which there is a transition from the decreasing to constant regime in the eigenvector average ratios. Sprinklings up to 2200 elements were considered in a diamond with side length $2L=1$. The results fit $\lambda_{transition} = (0.13 \pm 0.03) N^{(0.46 \pm 0.04)}$, represented in blue. For comparison, the $\sqrt{N}/(4 \pi)$ truncation curve is also included, represented in orange. }
\label{magnitude}
\end{figure}

We consider several collections of sprinklings, each collection with a different density, and apply the procedure laid out in the previous subsection to each collection. We then approximate the transitions from the decreasing to the constant regimes\footnote{The approximation is done via the intersection of a linear and constant fit, as shown in Figure \ref{averages}.} and record the eigenvalue magnitude, $\lambda_{transition}$, at which this occurs. Our results for $\lambda_{transition}$ are shown in Figure \ref{magnitude} versus the number of elements $N$ in their respective sets. We considered  sprinklings up to $2200$ elements and the best fit power law for our results was $\lambda_{transition} = (0.13 \pm 0.03) N^{(0.46 \pm 0.04)}$, noting that the errors given should be understood as lower bounds since there is an uncertainty going into the estimation of the two regimes of Figure \ref{averages}. 

\begin{figure}[h!]
\centering
\includegraphics[scale=0.8]{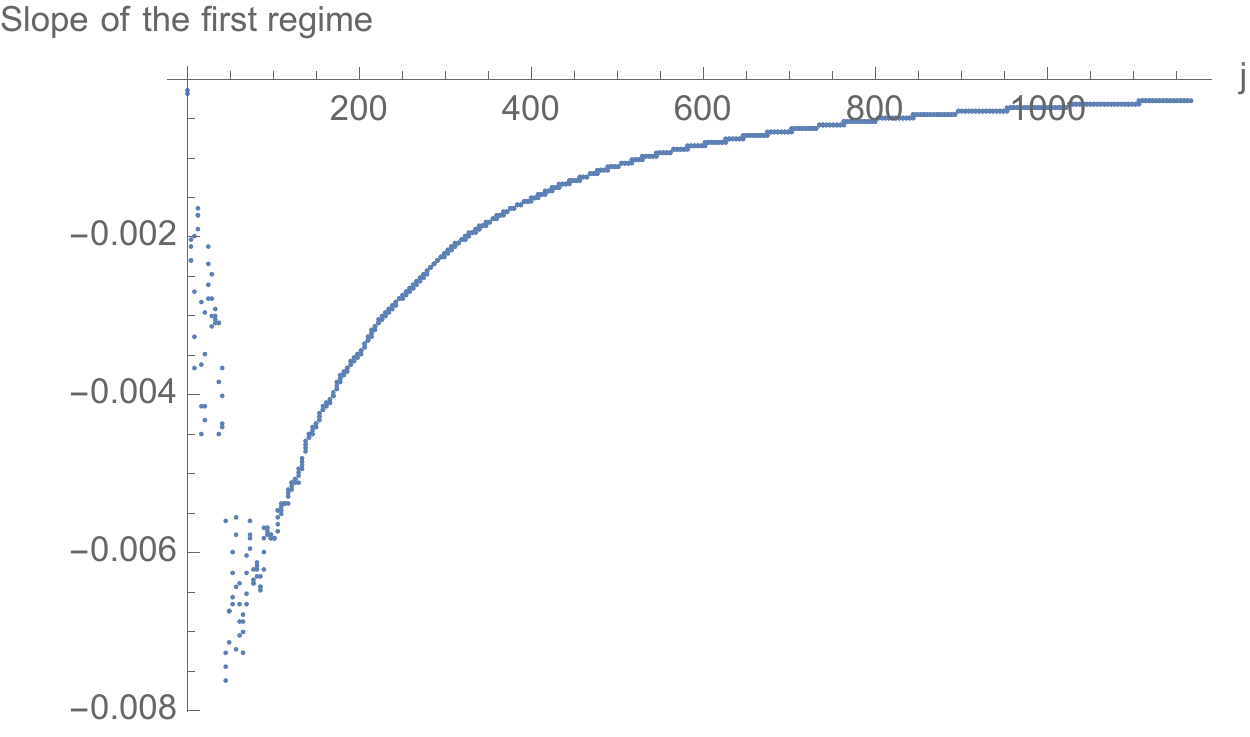}
\caption{Example plot of the slope of the linear fit representing the first regime as a function of the number of eigenvectors taken into account for N=1300.}
\label{slope}
\end{figure}

In order to reduce this uncertainty, the slope of the first regime was plotted against the number of elements, as shown in Figure \ref{slope}. As can be seen, a relatively clear minimum is present which removes some of the error in the distinguishing of the linear and constant regimes. The transition point was then decided by taking the average of these minima.

 The results for $\lambda_{transition}$ are close to our usual truncation rule in \eqref{mineig}. This gives evidence to the suggestion that there is a connection between the truncation and the transition to fluctuation-like eigenvectors. We have also checked that our results persist if we consider slightly different $N_c$ and $n$ values.

The connection between the transition to fluctuations and the truncation point is practically useful, as it provides an independent truncation scheme (using the eigenvectors) which generalises to any causal set (i.e. any dimension and with or without curvature). This is especially significant in cases where we lack a physical understanding of the eigenvalues and eigenfunctions of $i\Delta$, for example in terms of wavenumbers and plane waves, to reveal their relation to the discreteness scale.

\section{Discussion and Conclusions}
\label{summary}
In this work we extended previous work on the entanglement entropy of a massless scalar field theory on a $1+1$D causal set to R\'enyi entanglement entropies and the massive scalar field theory. In all the cases that we studied, we obtained results consistent with the expected  area law scalings (log scalings in $1+1$D). We used a spacetime definition of entropy and the covariant causal set discreteness scale as our UV cutoff. We also arrived at two key insights that will facilitate future generalisations of these studies to higher dimensions and causal sets approximated by curved spacetimes. 

The first insight was in our study of the massive theory, where we observed that the spectrum of the Pauli-Jordan function in the massive theory (where $m^2\ll\rho$) approaches the spectrum in the massless theory as we go deeper in the UV, towards the discreteness scale. The significance of this is that our knowledge of truncations, which are necessary to obtain meaningful entanglement entropies, can be carried over from the well studied massless theory to the less understood massive one. These truncations need to be implemented near the discreteness scale, which is precisely when the massless and massive spectra resemble one another. A conjecture is that this universality is more general. Namely, that the spectra in a different shaped compact region, in the presence of mass, curvature, and/or interactions, all approach that of the massless theory in the causal diamond, provided that all these other scales (mass, curvature, interaction strength, and system size) are far from the discreteness scale. This would mean that a general understanding of the spectrum of the spacetime commutator or Pauli-Jordan function in the UV would arise from our understanding of it in causal sets approximated by flat spacetime in every dimension of interest. 

The second novel insight that we obtained was regarding the nature of the truncated eigenvectors as fluctuations. We considered a particular averaging of the eigenvectors in an ensemble of coarse grained causal sets, and found evidence that the truncated (or discarded) eigenvectors have few features that persist after the averaging, whereas the non-truncated eigenvectors have large-scale/smooth variations and persistent features after the averaging procedure. This indicates that these components behave as fluctuations particular to individual causal sets. The physical significance of this requires further thought. Could this be a consequence of treating the causal set as a fixed background? Which quantities have a physical reality in every causal set realisation, and which quantities should only be regarded as physical if they persist over an ensemble of sprinklings? A full answer to these questions will require further development of causal set theory as a theory of quantum gravity. In this particular context of entanglement entropy, since we would like to recover the conventional area laws and study their relation to black hole entropy, it is the case that we need to exclude these fluctuations from our calculations. We have shown in this work how this can more generally be done.

\newpage

\bf Acknowledgements: \rm
We thank Fay Dowker for helpful comments. YY acknowledges financial support from Imperial College London through an Imperial College Research Fellowship grant.
\appendix
\section{Spacetime Entropy Derivation}\label{app:replica}
The entropy of the field theory can be broken up into a sum over the entropies of single degrees of freedom. Let us therefore consider a single degree of freedom with a conjugate pair of variables $q$ and $p$ satisfying $[q,p] = i$. The most general Gaussian density matrix, in the basis of $q$, is
\begin{equation}\label{eq:2.19}
\rho(q, q')\equiv \braket{q|\rho|q'} = \sqrt{\frac{A}{\pi}} e^{-\frac{A}{2}(q^2+q'^2)+i\frac{B}{2}(q^2-q'^2)-\frac{C}{2}(q-q')^2}
\end{equation}
where $A$, $B$, $C$ are constant parameters, and the normalisation constant $\sqrt{\frac{A}{\pi}}$ is fixed by the condition Tr$\rho=1$.

The replica trick is defined as \cite{Holzhey_1994}

\begin{equation}\label{replica}
    S= -\lim_{\alpha\to1} \frac{\partial}{\partial\alpha}\text{Tr}\,(\rho ^\alpha) \end{equation}

Following \cite{68} and computing this for our Gaussian density matrix \eqref{eq:2.19} we obtain 

\begin{flalign}{\label{trace}}
\begin{split}
&\text{Tr}\,(\rho^\alpha)=\bigg(\sqrt{\frac{A}{\pi}}\bigg)^\alpha \int dq_1 ... dq_\alpha\, \rho(q_1,q_2)\rho(q_2,q_3)...\rho(q_\alpha,q_1)\\
&=\bigg(\frac{A}{\pi}\bigg)^{\alpha/2}\int d^\alpha q\, \text{exp} \bigg(-(A+C)\sum_{n=1}^\alpha q_n^2 + C \sum_{n=1}^\alpha q_n q_{n+1} \bigg)=\frac{|1-\mu|^\alpha}{|1-\mu^\alpha|},
\end{split}
\end{flalign}
where $\mu=\frac{\sqrt{1+2C/A}-1}{\sqrt{1+2C/A}+1}$. Substituting this into \eqref{replica} then gives
\begin{flalign}{\label{finalvon}}
\begin{split}
&S=
-\lim_{\alpha\to1} \frac{\partial}{\partial\alpha}\bigg(\frac{|1-\mu|^\alpha}{|1-\mu^\alpha|}\bigg)
=-\frac{\mu \text{ln}\mu+(1-\mu)\text{ln}(1-\mu)}{1-\mu}.
\end{split}
\end{flalign}

Now let us relate this to $W$ and $i \Delta$ that appear in \eqref{eigeigeig}. For our single degree of freedom we have:
\begin{equation}\label{eq:2.20}
i\Delta = 2\,\text{Im}\bigg(\begin{matrix}\braket{qq}
 &\braket{qp} \\ \braket{pq}
 & \braket{pp}
\end{matrix}\bigg)= \bigg(\begin{matrix}0
 &1 \\ -1
 & 0
\end{matrix}\bigg)
\end{equation}
\begin{equation}\label{eq:2.21}
W = \bigg(\begin{matrix}\braket{qq}
 &\braket{qp} \\ \braket{pq}
 & \braket{pp}
\end{matrix}\bigg)=\bigg(\begin{matrix}1/(2A)
 &i/2 \\ -i/2
 & A/2+C
\end{matrix}\bigg)
\end{equation}

The eigenvalues of $(i\Delta)^{-1}W$ which enter the entropy definition \eqref{see} are $\lambda_+=1/2+\sqrt{1/4+C/(2A)}$ and $\lambda_-=1-\lambda_+$, such that

\begin{equation}
    S=\lambda_+\ln\lambda_++\lambda_-\ln|\lambda_-|
\end{equation}

Rewriting this in terms of $\mu$ we arrive at the same expression as in \eqref{finalvon}. Hence the entropy \eqref{see} agrees with the one in \eqref{finalvon}. Passing over to the full field theory, we sum over all eigenvalues of $(i\Delta)^{-1}W$ to arrive at the  entropy \eqref{see}.

Similarly, the formulas for R\'enyi entropies in terms of the eigenvalues of $(i\Delta)^{-1}W$ can be shown to be

\begin{flalign}{\label{finalren}}
\begin{split}
&S^{(\alpha)}(\rho) = \frac{1}{1-\alpha}\text{ln}\,\text{Tr} (\rho^\alpha)=\frac{1}{1-\alpha}\text{ln}\bigg(\frac{|1-\mu|^\alpha}{|1-\mu^\alpha|}\bigg)\\
&=\frac{1}{1-\alpha}\text{ln}\bigg(\frac{(1/\lambda_+)^\alpha}{(\lambda_+^\alpha-(\lambda_+-1)^\alpha)/\lambda_+^\alpha}\bigg)\\
&=-\frac{1}{1-\alpha}\text{ln}(\lambda_+^\alpha-(\lambda_+-1)^\alpha).
\end{split}
\end{flalign}
 Summing over the full spectrum of $(i\Delta)^{-1}W$ gives the full field theory Rényi entropy of order $\alpha$ as the sum
\begin{equation}\label{ftra}
    S^{(\alpha)}=\frac{-1}{1-\alpha}\sum_\lambda \text{ln}(\lambda^\alpha - (\lambda-1)^\alpha),
\end{equation}
where each term in the sum \eqref{ftra} accounts for a pair of eigenvalues $\lambda$ and $(1-\lambda)$.

\section*{References}
\bibliographystyle{ieeetr}
\bibliography{main}
\end{document}